\documentclass[journal,twoside,web]{ieeecolor}
\usepackage{generic}
\usepackage{cite}
\usepackage{amsmath,amssymb,amsfonts}
\usepackage{algorithmic}
\usepackage{graphicx}
\usepackage{algorithm,algorithmic}
\usepackage{textcomp}
\usepackage{mathrsfs}
\usepackage{subcaption}

\def\BibTeX{{\rm B\kern-.05em{\sc i\kern-.025em b}\kern-.08em
    T\kern-.1667em\lower.7ex\hbox{E}\kern-.125emX}}
\newtheorem{theorem}{Theorem}

\newtheorem{lemma}{Lemma}

\newtheorem{definition}{Definition}
\newtheorem{assumption}{Assumption}
\newtheorem{remark}{Remark}
\newenvironment{lemwithref}[2]
  {\renewcommand\thelemma{#1 \cite{#2}}\lemma}
  {\endlemma}

\usepackage{hyperref}
\usepackage{fancyhdr}
\fancypagestyle{arxiv}{%
  \fancyhf{} % 清空默认
  \fancyhead[C]{%
    This is the author’s version of the article accepted for publication in \textit{IEEE Transactions on Automatic Control}%
  }
  \fancyfoot[C]{%
    Final version available at \href{https://ieeexplore.ieee.org/abstract/document/11184745}{doi:10.1109/TAC.2025.3616202}%
  }
}

\begin{document}
\title{Distributed Feedback-Feedforward Algorithms for Time-Varying Resource Allocation}
\author{Yiqiao Xu, \IEEEmembership{Member, IEEE}, Tengyang Gong, \IEEEmembership{Graduate Student Member, IEEE}, Zhengtao Ding, \IEEEmembership{Senior Member, IEEE}, and Alessandra Parisio, \IEEEmembership{Senior Member, IEEE}
\thanks{This work was partially supported by EPSRC projects: Supergen Energy Networks Impact Hub (EP/Y016114/1), Multi-energy Control of Cyber-Physical Urban Energy Systems (EP/T021969/1), and Grid Scale Thermal and Thermo-Chemical Electricity Storage (EP/W027860/1).}
\thanks{Yiqiao Xu, Tengyang Gong, Zhengtao Ding, and Alessandra Parisio are with the Department of Electrical and Electronic Engineering, The University of Manchester, Manchester, M13 9PL U.K. (e-mail: yiqiao.xu; tengyang.gong; zhengtao.ding; alessandra.parisio@manchester.ac.uk).}
\thanks{\noindent
\copyright~2025 IEEE. Personal use of this material is permitted. 
Permission from IEEE must be obtained for all other uses, in any current or future media, 
including reprinting/republishing this material for advertising or promotional purposes, 
creating new collective works, for resale or redistribution to servers or lists, 
or reuse of any copyrighted component of this work in other works.  
The final version of record is available at:  
\href{https://ieeexplore.ieee.org/abstract/document/11184745}{10.1109/TAC.2025.3616202}.
}
}

\maketitle
\thispagestyle{arxiv}   % 首页页脚加声明
\pagestyle{arxiv}       % 后续页也加声明
\begin{abstract}
    This paper studies distributed Time-Varying Resource Allocation (TVRA) where the local cost functions, global equality constraints, and Local Feasibility Constraints (LFCs) vary with time. Algorithms that mimic the structure of feedback-feedforward control systems are proposed. Feedback and feedforward laws are generated using local estimates from a distributed estimator, while a distributed controller enforces the stationarity condition within a fixed time and updates the candidate solution accordingly. To handle the LFCs, feedback laws based on projection and feedforward laws that switch between different modes are introduced as an initialization-free alternative to the barrier-based methods used in most related works. Our projection-based method guarantees that, for any infeasible initial value, the state trajectory enters the locally feasible set within a fixed time and remains within it thereafter, and that the set is forward invariant if the initial value is locally feasible. Convergence analyses are conducted under mild assumptions. For cases without LFCs, the proposed algorithm converges to the optimal trajectory within a fixed time. For cases with LFCs, the proposed algorithm is globally asymptotically stable at the optimal trajectory while exhibiting fixed-time convergence between consecutive switching instants. Numerical examples and a power system application verify their effectiveness.
\end{abstract}

\section{Introduction}
\label{sec:introduction}

\IEEEPARstart{R}{ecently}, Time-Varying Optimization (TVO) \cite{7469393} has gained increasing attention in both academia and industry, with applications in energy conversion systems \cite{8782575}, smart grids \cite{9810516,9133310}, traffic engineering \cite{su2009traffic}, vehicular systems \cite{9133310,9855238}, and quadrotors \cite{9670443}. The need to accommodate time-varying cost functions and/or constraints necessitates a shift from static to dynamic optimization algorithms to find and track the optimal trajectory under evolving conditions. As a subclass of TVO, Time-Varying Resource Allocation (TVRA) imposes more challenging constraints and stricter feasibility requirements, since the resources allocated to activities are physically limited, manifesting hard constraints. The challenge becomes even more pronounced in multi-agent systems, where agents have access only to partial problem information and must coordinate via distributed communication protocols. Compared to time-invariant optimization, distributed TVO and especially distributed TVRA remain underexplored.

\subsection{Related works}
The research on TVO can be traced back to preliminary results known as the ``prediction-correction method" \cite{7469393, 7993088, 8502778}, where problems are sampled at discrete time steps and solved online. At each time step, newly sampled data are used to correct the candidate solution and to predict the evolution of the optimal trajectory for the next step. A continuous-time variant of this method is proposed in \cite{8062794}, which tracks the optimal trajectory of a centralized TVRA problem with asymptotically vanishing error. However, its application to multi-agent systems is hindered because the prediction step requires computing the inverse of the Hessian matrix, which represents global information. Although a decentralized approximation of the Hessian inverse can be obtained by truncating its Taylor expansion, this approach relies on augmented Lagrangian relaxation and inevitably introduces an optimality gap between the relaxed and original problems \cite{7902101}.

There are only a few works in the literature that address distributed TVO problems, most of which are characterized by state consensus constraints. In \cite{7862771}, it is shown that, in the absence of Local Feasibility Constraints (LFCs), consensus-constrained distributed TVO problems are equivalent to unconstrained ones, which are more amenable to being solved. In the same work, although a projection-based gradient algorithm is used when LFCs are present, tracking errors with respect to the optimal trajectory persist. In \cite{8100702}, a fixed-time consensus protocol is developed to guarantee an upper bound on the settling time for state consensus under the assumption of identical Hessians. The important role of distributed estimator \cite{7518617} in handling non-identical Hessians has been well acknowledged, as evidenced in works such as \cite{9629369, chen2023time, li2020time, 9827560, zhao2025distributed}. In particular, fixed-time distributed estimators are designed in \cite{li2020time,9827560} to estimate the time derivative of the optimal trajectory, thereby enabling vanishing tracking errors. Distributed TVRA faces a class of less tractable constraints, including global equality/inequality constraints and LFCs. In recent years, several distributed algorithms for TVRA have been proposed, assuming uniformly strongly convex local cost functions: 1) with time-invariant Hessians \cite{9183902}, 2) with identical time-varying Hessians \cite{8619295, 9123943}, and 3) with non-identical time-varying Hessians \cite{9670443, 10239516,9855238,10791871}. In \cite{9953144}, the author shows that TVO problems (including distributed TVRA problems) can be recast as output regulation problems \cite{isidori1990output} and solved using algorithms motivated by the feedforward control design, albeit limited to periodic temporal variations. To tackle aperiodic temporal variations, signum-based consensus protocols such as finite-time and fixed-time consensus protocols are generally required.

LFCs are often omitted in related works (\textit{e.g.}, \cite{9670443,9893893,9183902,9123943,10239516,9953144,10791871}) despite their prevalence in practical applications. The inclusion of LFCs will cause the optimal trajectory to evolve differently on the boundary and in the interior of the locally feasible set, rendering the above results not directly applicable. To the best of our knowledge, methods effectively addressing LFCs have so far been exclusively barrier-based \cite{chen2023time, 9855238, 10829801, zhao2025distributed}, tracing their origins to the ``prediction-correction interior-point method'' \cite{8062794}. The core idea is to use a time-dependent log-barrier function to confine the state trajectory within the locally feasible set, and to relax initialization requirements by introducing a time-dependent slack variable that temporarily enlarges the set. A challenge associated with this time dependence is the trade-off between feasibility and initialization, as well as the difficulty in implementing finite/fixed-time distributed estimators \cite{7518617, 9629369, chen2023time, li2020time, 9827560, zhao2025distributed} due to the monotonically increasing barrier parameters. While dualization \cite{8746216} and interior-point methods \cite{boyd2004convex} have long served as foundational tools for constraint handling in distributed optimization, recent literature on time-invariant problems has explored several other techniques such as proximal point \cite{9762539,huang2024distributed}, projection \cite{9141512,yi2016initialization,liu2024achieving,10909192}, and control barrier functions \cite{10876582}. These results show violation-bounded \cite{9762539,huang2024distributed,10909192} or even violation-free performance \cite{9141512,yi2016initialization,liu2024achieving,10876582} in time-invariant settings but do not address the time-varying nature of distributed TVRA, which motivates our work.

\subsection{Contributions}
Log-barrier functions remain the predominant tool for handling LFCs in distributed TVO \cite{chen2023time, 9855238, 10829801, zhao2025distributed}, as the use of projection \cite{7862771} and penalty functions \cite{8619295} typically introduces an optimality gap, underscoring the scarcity of effective alternatives. This paper aims to explore an alternative to the barrier-based methods for distributed TVRA problems. The studied problems feature uniformly strongly convex local cost functions, global affine equality constraints, and uniformly convex local feasibility constraints, all of which are time-varying. Fully distributed algorithms that mimic the structure of feedback-feedforward control systems are proposed for cases without and with LFCs. Given non-identical time-varying Hessians, both algorithms consist of two parts: a distributed estimator that generates local estimates for the feedback and feedforward laws, and a distributed controller that enforces the stationarity condition within a fixed time and updates the candidate solution accordingly. Particularly, feedback laws based on projection and feedforward laws that switch between different modes are devised to handle the LFCs. A state-dependent switching signal that synchronizes with projection is devised to govern the feedforward laws and reduce the number of switches without compromising optimality. Convergence analyses are conducted under mild assumptions. For cases without LFCs, the proposed algorithm converges to the optimal trajectory within a fixed time. For cases with LFCs, the proposed algorithm is globally asymptotically stable at the optimal trajectory while exhibiting fixed-time convergence between consecutive switching instants. Unlike the barrier-based methods \cite{chen2023time,8062794, 9855238, zhao2025distributed,10829801} that need to be well initialized and/or track an approximate optimal trajectory until the barrier parameters approach infinity, our projection-based method is initialization-free and tracks the exact optimal trajectory. The state trajectory remains within the locally feasible set if initialized therein; when started from an arbitrary infeasible initial value, the state trajectory is guaranteed to enter the locally feasible set within a fixed time and stay there subsequently. This initialization-free capability, coupled with a guarantee of local feasibility, is very convenient in many practical applications. Practically, it achieves global fixed-time convergence to the optimal trajectory, with negligible effects from switching. Additionally, we show that switching is non-Zeno and that the projection-based method can be extended to handle global affine inequality constraints. Numerical examples and a power system application (\textit{i.e.}, a new fast frequency response service in the U.K.) are presented to verify the effectiveness of the proposed algorithms.

The remainder of this paper is organized as follows. Section \uppercase\expandafter{\romannumeral2} introduces preliminaries and mathematical background. Section \uppercase\expandafter{\romannumeral3} presents the studied problems and the main results. Section \uppercase\expandafter{\romannumeral4} verifies the effectiveness of the proposed algorithms through case studies. Section \uppercase\expandafter{\romannumeral5} concludes this paper. Finally, Section \uppercase\expandafter{\romannumeral6} contains the appendix.

\section{Preliminaries}
\subsection{Notations and Graph Theory}
\begin{table}[b]
    \centering
    \caption{List of key notations for each agent.}
    \label{table}
    \begin{tabular}{|p{55pt}| p{165pt}| }
        \hline
        $A_i$, $b_i$  &  Allocation matrix and local activity function\\
        \hline
        $x_i$, $x_i^\star$ & decision variable and its optimal trajectory\\
        \hline
        $\lambda_i$, $\lambda^\star$  & dual variable and its optimal trajectory\\
        \hline
         $f_i$   &  local cost function\\
        \hline
        $f_{ix}$  & partial derivative of $f_i$ with respect to $x_i$\\
        \hline
        $f_{ixx}$, $f_{ixt}$  &  partial derivatives of $f_{ix}$ with respect to $x_i$ and $t$\\
        \hline
        $b_{it}$  &  partial derivative of $b_{i}$ with respect to $t$\\
        \hline
        $F_{i,x}$, $F_{i,\lambda}$, $F'_{i,x}$ & feedback laws\\
        \hline
        $\alpha_{i,x}$, $\alpha_{i,\lambda}$, $\alpha'_{i,x}$ & feedforward laws\\
        \hline
        $y_{i}$, $\psi_i$, $\psi'_i$, $\Delta_i$ & local estimates\\
        \hline
        $\theta_{i}$, $\theta'_{i}$, $\zeta_{i}$ & intermediate variables\\
        \hline
        $\rho_{i}$, $\phi_{i}$, $G_i$, $H_i$ & weight matrix/vector\\
        \hline
        $e_i$ & error with respect to the stationarity condition\\
        \hline
        $g_{i,m}$  &  local feasibility constraint\\
        \hline
        $\sigma_{i,m}$ & state-dependent switching signal\\
        \hline
        $X_{i}$ &  locally feasible set\\
        \hline
    \end{tabular}
\end{table}
The following notations are used throughout this paper. The sets of real numbers, non‑negative real numbers, and positive real numbers are denoted by $\mathbb R$, $\mathbb R_{+}$, and $\mathbb R_{++}$, respectively. The sets of integers, non-negative integers, and positive integers are denoted by $\mathbb{Z}$, $\mathbb{Z}{+}$, and $\mathbb{Z}{++}$, respectively. $\mathbb R^n$ denotes the $n$‑dimensional real coordinate space, and $\mathbb R^{m\times n}$ denotes the set of all real $m \times n$ matrices. We use $\operatorname{sign}(\cdot)$, $\Vert\cdot\Vert$, $\Vert\cdot\Vert_1$, $\Vert\cdot\Vert_2$, and $\Vert\cdot\Vert_p$ to denote the component-wise signum, absolute value, $\ell_1$‑norm, $\ell_2$‑norm, and $\ell_p$‑norm of $(\cdot)$, respectively. Moreover, $(\cdot)^\top$, $(\cdot)^{-1}$, $(\cdot)^\dagger$, and $\circ$ represent transpose, inverse, pseudoinverse, and element-wise multiplication, respectively. The communication network among $N$ agents (labeled $1$ through $N$) is modeled as a directed, unweighted graph $\mathcal{G}=(\mathcal{N},\mathcal{E})$, where $\mathcal{N}=\{1,2,\dots,N\}$ is the set of vertices and $\mathcal{E}\subseteq\mathcal{N}\times\mathcal{N}$ is the set of edges. Each vertex $i\in\mathcal{N}$ corresponds to agent $i$, and the existence of a directed edge $(i,j)\in\mathcal{E}$ indicates that agent $i$ can receive information from agent $j$. The adjacency matrix $A=[a_{ij}]\in\mathbb R^{N\times N}$ is defined by $a_{ij}=1$ if $(i,j)\in\mathcal E$ and $a_{ij}=0$ otherwise. If $(i,j)\in\mathcal{E}$ and also $(j,i)\in\mathcal{E}$, we say the edge between $i$ and $j$ is undirected. Define the Laplacian matrix $\mathcal L=[l_{ij}]_{N\times N}$ for the graph $\mathcal G$ by $l_{ii}=\sum_{j=1}^Na_{ij}$ and $l_{ij}=-a_{ij}$ whenever $j\neq i$. The graph $\mathcal{G}$ is called connected if for every pair of distinct vertices $i,j\in\mathcal{N}$, there exists a spanning tree between them. It is called undirected if every edge is bidirectional, \textit{i.e.}, $(i,j)\in\mathcal{E}\iff(j,i)\in\mathcal{E}$.  For an undirected graph, zero is a simple eigenvalue of $\mathcal{L}$ if and only if $\mathcal{G}$ is connected. We denote the second smallest eigenvalue of $\mathcal{L}$ by $\eta_2(\mathcal{L})$.

\subsection{Convex and Nonsmooth Analyses}
\begin{definition}
(Convexity)  
Let $X\subset\mathbb{R}^n$ be a convex set and let $f:X\to\mathbb{R}$. Then, $f$ is called convex on $X$ if, for every $x,y\in X$ and every $\eta\in[0,1]$, one has
$f(\eta x + (1-\eta)y)\leq\eta f(x)+(1-\eta)f(y)$. Equivalently, if $f$ is differentiable on $X$, then $f$ is convex if and only if for all $x,y\in X$, $\langle f(y)-\nabla f(x),y-x\rangle \leq0$. Furthermore, $f$ is said to be $m$-strongly convex on $X$ (for some constant $m>0$) if, for every $x,y\in X$, $f(y)\geq f(x)+\langle \nabla f(x),y-x\rangle +\frac{m}{2}\Vert y-x\Vert_2^2$. Equivalently, if $f$ is twice continuously differentiable on $X$, then $f$ is $m$-strongly convex if and only if $\nabla^2 f(x) \succeq\ mI_n$ $\forall x\in X$.
\end{definition}

\begin{definition}
(Projection) Let $X \subset \mathbb R^n$ be a nonempty, closed, convex set. For any $x\in\mathbb R^n$, define the projection of $x$ onto $X$ by $\mathcal P_{X}(x)$, which is the unique point in $X$ satisfying
\begin{align}
    \mathcal P_{X}(x) = \mathrm{arg}\min\nolimits_{y\in X}\Vert y-x\Vert_2^2.
\end{align}
The normal cone of $X$ at any $x \in X$ is given by
\begin{align}\label{cone}
    \mathcal C_X(x) = \{v\in\mathbb R^n\mid \mathcal P_{X}(x+v)=x\}.
\end{align}
\end{definition}

\begin{definition}
(Filippov Solution) An absolutely continuous mapping $x:[0,T]\to\mathbb R^n$ is a Filippov solution of the discontinuous Ordinary Differential Equation (ODE) $\dot x(t)=f(x(t))$, $x(0)=x_0$, if for almost all $t\in[0,T]$, $\dot x(t)\in K[f](x(t))$. Equivalently, one regards the original ODE as the differential inclusion $\dot x(t)\in K[f](x(t))$.
\end{definition}
\begin{definition}
(Chain Rule) Let $f:\mathbb R^n\to\mathbb R^n$ be essentially bounded on compact sets, and let $K[f]$ be its Filippov set‐valued map. Suppose $x:[0,T]\to\mathbb R^n$ is a Filippov solution of $\dot x(t)=f(x(t))$. Let $V:\mathbb R^n\to\mathbb R$ be locally Lipschitz, and denote its Clarke generalized gradient by $\partial V(x(t))$. Then for almost all $t\in[0,T]$, $\dot V(x(t))\in\dot{\tilde V}(x(t))$, where $\dot{\tilde V}(x(t))=\{\langle a,b\rangle \mid a\in\partial V(x(t)), b\in K[f](x(t))\}$ is the set-valued Lie derivative of $V(x(t))$.
\end{definition}

\subsection{Fixed-Time Stability}
\begin{lemwithref}{1}{8322314}\label{lemma zuo}
Consider a dynamical system $\dot x(t) = f(x(t))$, $x(0)=x_0$, where $x(t) \in \mathbb R^n$. Suppose there exists a continuous, positive definite, radially unbounded Lyapunov function $V: \mathbb{R}^n \to \mathbb{R}_+$ such that $V(x(t))=0$ if $x(t)=\mathbf 0_n$, and for all $x(t) \neq \mathbf 0_n$,
\begin{align}\label{differential equation}
        \dot V(x(t)) \leq -\gamma_{1} V(x(t))^{1-\frac{p}{q}} - \gamma_{2} V(x(t))^{1+\frac{p}{q}},
    \end{align}
where $\gamma_1,\gamma_2\in\mathbb{R}_{++}$, $p$ is an even integer, and $q$ is an odd integer satisfying $0<p<q$. Then the origin is fixed-time stable, and the settling time $T$ satisfies the upper bound:
\begin{align}
    T\leq T^{\max}=\frac{\pi q}{2p\sqrt{\gamma_{1}\gamma_{2}}},\ \forall x_0\in\mathbb R^n.
\end{align}
\end{lemwithref}

\begin{definition}
(Fixed-Time Consensus)  
Consider a dynamical system with state variable $x(t) \in \mathbb{R}^N$. Then $x(t)$ is said to achieve fixed-time consensus if there exists a constant $T^{\max}\in\mathbb{R}_{++}$ such that the following conditions are satisfied for all $i,j\in\mathcal{N}$:
\begin{subequations}
\begin{align}
    \lim_{t \to T} \Vert x_i(t) - x_j(t) \Vert_2 &= 0,\\
    \Vert x_i(t) - x_j(t) \Vert_2 &= 0,\ \forall t \geq T,
\end{align}
\end{subequations}
where $T\in [0,T^{\max}]$ for any initial value $x(0)$.
\end{definition}

\section{Main Results}
Many engineering problems can be formulated as resource allocation problems, which become time-varying if the cost functions and/or constraints are parameterized by time. This section presents our results on distributed TVRA, starting with the case without LFCs and followed by the case with LFCs.

\subsection{Feedback-Feedforward Algorithm without LFCs}
Consider a distributed TVRA problem in which both the cost functions and the global equality constraints vary with time. Each agent holds a local cost function and partial knowledge of the global equality constraints. The goal is to find and track the optimal trajectory $x^\star(t)$ that minimizes the total cost and allocates resources to activities, using only local information and communication with neighboring agents:
\begin{align}\label{problem}
\begin{split}
    x^\star(t) &= \underset{x(t)\in\mathbb R^{Nn}}{\mathrm{arg}\min}\sum_{i=1}^N f_i(x_i(t),t),\\
    \text{s.t. }&\sum_{i=1}^N A_ix_i(t) = \sum_{i=1}^N b_i(t).   
\end{split}
\end{align}
For agent $i$, $x_i(t)\in\mathbb R^n$ represents its resource to allocate. The local cost function is denoted by $f_i(x_i(t),t):\mathbb R^n\times\mathbb R_{+}\to\mathbb R$, and the local activity function is given by $b_i(t):\mathbb R_{+}\to\mathbb R^l$. The allocation matrix $A_i\in\mathbb R^{l\times n}$ maps resources to activities, resulting in $l$ global equality constraints.

\begin{assumption}\label{assumption1}
The local cost function $f_i(x_i,t)$ is uniformly strongly convex in $x_i$ and twice continuously differentiable with respect to $t$ for all $t\geq 0$ and $i\in\mathcal N$. The local activity function $b_i(t)$ is continuously differentiable with respect to $t$ for all $t\geq 0$ and $i\in\mathcal N$. Moreover, $A_i$ has full rank, and the number of global equality constraints satisfies $l\leq n$.
\end{assumption} 
\begin{assumption}\label{assumption2}
The communication graph $\mathcal G$ is connected, undirected, and fixed for all $t\geq 0$.
\end{assumption}

Let $\bar\lambda\in\mathbb R^l$ denote the Lagrange multiplier associated with the global equality constraints. For distributed optimization, $\bar\lambda$ is replaced by local dual variables $\lambda_i\in\mathbb R^l$, and it is required that $\lambda_i\to\bar\lambda$ for all $i\in\mathcal N$, where $\bar\lambda = \frac{1}{N}\sum_{i=1}^N\lambda_i$. Define the vector of primal-dual variables for agent $i$ as $z_i=\text{col}(x_i,\lambda_i)\in\mathbb R^{n+l}$. As shown in \cite{9953144}, which, however, targets periodic temporal variations, primal-dual dynamics of the following form could be constructed to solve problem (\ref{problem}):
\begin{align}\label{of the following form}
    \dot z_i &= - F_i(z_i,t) + \alpha_i(z_i,y_i,t) -\text{col}(\mathbf 0_n,\mathscr{C}_i(\lambda)),
\end{align}
where $F_i(z_i,t):\mathbb R^{n+l}\times\mathbb R_{+}\to\mathbb R^{n+l}$ summarizes the feedback laws, $\alpha_i(z_i,y_i,t):\mathbb R^{n+l}\times\mathbb R^l\times\mathbb R_{+}\to\mathbb R^{n+l}$ summarizes the feedforward laws, and $y_i\in\mathbb R^l$ is a local estimate of $\dot{\bar\lambda}$. With a proper consensus protocol $\mathscr{C}_{i}(\lambda)$,
\begin{align}\label{dynamical system optimal}
\dot z_i^\star &= -F_i(z_i^\star,t) + \alpha_i(z_i^\star,y_i^\star,t),
\end{align}
where $z_i^\star = \text{col}(x_i^\star,\lambda^\star)$ represents the optimal trajectory of $x_i$ and $\bar\lambda$, and $y_i^\star=\lim_{z_i\to z_i^\star}y_i$.

The basic concept of feedforward control is to estimate and compensate for the temporal variations before they upset tracking of the optimal trajectory:
\begin{align}
\alpha_i(z_i^\star,y_i^\star,t) = \dot z_i^\star.
\end{align}
Correspondingly, the feedback laws should be designed in a way such that
\begin{align}\label{the following condition is consistently satisfied}
F_i(z_i^\star,t) = \mathbf 0_{n+l},
\end{align}
which motivates our use of fixed-time control to enforce $F_i(z_i,t)$ to the origin. In the absence of LFCs, $\dot z_i^\star$ can be directly comprehended from the problem formulation.

\begin{lemma}\label{lemma optimal trajectory}
Suppose Assumption 1 hold. Define
\begin{subequations}\label{weight}
\begin{align}
    \rho_i &= A_if_{ixx}(x_i,t)^{-1}A_i^\top,\\
    \phi_i &=A_if_{ixx}(x_i,t)^{-1}f_{ixt}(x_i,t)+b_{it}(t),
\end{align} 
\end{subequations}
Let $x^\star$ and $\lambda^\star$ denote the optimal trajectory of primal and dual variables for problem (\ref{problem}). Then,
\begin{subequations}\label{analytical expressions}
\begin{align}
    \dot x_i^\star &= -f_{ixx}(x_i^\star,t)^{-1}(A_i^\top\dot\lambda^\star+f_{ixt}(x_i^\star,t)),\\
    \dot\lambda^\star &= -\left[\sum_{i=1}^N\rho_i\right]^{-1}\sum_{i=1}^N\phi_i.
\end{align}
\end{subequations}
\end{lemma}

\begin{proof}
    See Appendix.A.
\end{proof}

To this end, we propose a feedback-feedforward algorithm composed of two parts to solve problem (\ref{problem}). The first part is a distributed estimator that gathers local estimates to compute the feedback and feedforward laws. The second part is a distributed controller that implements these laws and updates the primal-dual variables accordingly. The distributed estimator is as follows:
\begin{subequations}\label{distributed estimator}
\begin{align}
    y_i & = \psi_i^{\dagger}\psi'_i,\\  
    \dot\theta_i &= -\mathscr{C}_{i}(\psi),\ \theta_i(0) = \mathbf 0_{l\times l},\\
    \psi_i &= \theta_i + \rho_i,\\
    \dot\theta'_i &= -\mathscr{C}_{i}(\psi'),\ \theta'_i(0) = \mathbf 0_l,\\
    \psi'_i &= \theta'_i - \phi_i,\\
    \dot\zeta_i &= -\mathscr{C}_{i}(\Delta),\ \zeta_i(0) = \mathbf 0_l,\\
    \Delta_i &= \zeta_i + b_i(t)-A_ix_i,
\end{align} 
\end{subequations}
where $y_i\in\mathbb R^l$ will be used to compute the feedforward laws; $\psi_i\in\mathbb R^{l\times l}$, $\psi'_i\in\mathbb R^l$, $\Delta_i\in\mathbb R^l$ are local estimates; $\theta_i\in\mathbb R^{l\times l}$, $\theta'_i\in\mathbb R^l$, $\zeta_i\in\mathbb R^l$ are intermediate variables; $\rho_i\in\mathbb R^{l\times l}$, $\phi_i\in\mathbb R^l$ are time-varying weight matrix/vector defined in (\ref{weight}); $\mathscr{C}_{i}(\cdot)$ represents the consensus protocol for $(\cdot)$, given by $\mathscr{C}_{i}(\cdot) = \gamma_{1}\sum_{j=1}^Na_{ij}((\cdot)_i-(\cdot)_j)^{1-\frac{p}{q}} +\gamma_{2}\sum_{j=1}^Na_{ij}((\cdot)_i-(\cdot)_j)^{1+\frac{p}{q}} + \gamma_{3,(\cdot)}\sum_{j=1}^Na_{ij}\operatorname{sign}((\cdot)_i-(\cdot)_j)$ \cite{8100702}. Note that $\gamma_1,\gamma_2,\gamma_{3,e}\in\mathbb R_{++}$ are control parameters to select, $p$ is a positive even integer, $q$ is a positive odd integer satisfying $p<q$, and $\mathscr{C}_{i}(\lambda)$, $\mathscr{C}_{i}(\psi)$, $\mathscr{C}_{i}(\psi')$, and $\mathscr{C}_{i}(\Delta)$ share the same design but may have different values of $\gamma_{3,(\cdot)}$. 

\begin{remark}\label{remark singularity}
In (\ref{distributed estimator}), $\psi_i$ represents a local estimate of $\frac{1}{N}\sum_{i=N}\rho_i$, which is diagonal and positive definite. However, $\psi_i$ may become arbitrarily small or singular before converging to the desired value. To prevent this, $y_i$ should be set to $\mathbf 0_l$ when the determinant of $\psi_i$ falls below a certain threshold, or alternatively, projected onto a reasonable hyperrectangle. Without such measures, the proposed algorithms may suffer from degraded performance or even fail.
\end{remark}

\begin{assumption}\label{assumption3}
The control parameters are sufficiently large to satisfy $\gamma_{3,\psi}\geq\frac{N-1}{2}\Vert\dot\rho_i\Vert_2$, $\gamma_{3,\psi'}\geq\frac{N-1}{2}\Vert\dot\phi_i\Vert_2$, and $\gamma_{3,\Delta}\geq\frac{N-1}{2}\Vert A_i\dot x_i-b_{it}(t)\Vert_2$ $\forall t\geq 0$ $\forall i\in\mathcal N$.
\end{assumption}

Control laws are designed separately for the primal and dual variables. In relation to (\ref{of the following form}), let $\text{col}(F_{i,x},F_{i,\lambda}) = F_i(z_i,t)$ and $\text{col}(\alpha_{i,x},\alpha_{i,\lambda}) = \alpha_i(z_i,y_i,t)$. The distributed controller is as follows:
\begin{subequations}\label{distributed controller}
\begin{align}
    \dot x_i &= -F_{i,x} + \alpha_{i,x},\\
    \dot\lambda_i &= - F_{i,\lambda} + \alpha_{i,\lambda} - \mathscr C_i(\lambda),\\
    e_i &= f_{ix}(x_i,t) + A_i^\top\lambda_i,\\
    F_{i,x} &= f_{ixx}(x_i,t)^{-1}(\gamma_1e_i^{1-\frac{p}{q}}+\gamma_2e_i^{1 + \frac{p}{q}}+\gamma_{3,e}\operatorname{sign}(e_i)),\\
    F_{i,\lambda} &= \beta \Delta_i,\\
    \alpha_{i,x} &= -f_{ixx}(x_i,t)^{-1}(A_i^\top y_i+f_{ixt}(x_i,t)),\\
    \alpha_{i,\lambda} &= y_i,
\end{align}
\end{subequations}
where $e_i\in\mathbb R^n$ denotes the error with respect to the stationarity condition; $f_{ix}(x_i,t)$ represents the gradient; $f_{ixx}(x_i,t)$ and $f_{ixt}(x_i,t)$ represent the Hessians; and a relatively large $\beta\in\mathbb R_{++}$ is used in (\ref{distributed controller}e) to retain fixed-time convergence under chattering, as further discussed in Remark \ref{remark beta}.

\begin{remark}\label{remark filippov}
The ODEs in (\ref{distributed estimator})--(\ref{distributed controller}) have discontinuous right-hand sides. Solutions to such ODEs might not be unique, or might not exist in the classical sense. Strictly speaking, they should be described by differential inclusions, and convergence of such nonsmooth systems should be analyzed in the sense of Filippov solutions. Nevertheless, since the state trajectories (\textit{e.g.}, $\dot x_i$) are measurable and locally essentially bounded, Filippov solutions exist for all time. Furthermore, at points of discontinuity, the set-valued Lie derivative of any locally Lipschitz Lyapunov function reduces to a singleton \cite{shevitz1994lyapunov}. Considering that the proofs remain the same for almost all $t$, we omit differential inclusions in the proofs to avoid notational redundancy.
\end{remark}

Before proceeding to convergence analysis, we define the following quantities to facilitate our development:
\begin{align}
   T_1^{\max} &= \frac{\pi q l^\frac{p}{4q}N^\frac{p}{2q}}{2p\sqrt{\gamma_1\gamma_2}\eta_2(\mathcal L)},\label{T1}\\ 
   T_2^{\max} &= \frac{\pi q n^\frac{p}{4q}}{2p\sqrt{\gamma_1\gamma_2}}.\label{T2}
\end{align}

\begin{lemma}\label{lemma distributed estimator}
Consider problem (\ref{problem}) and the distributed estimator in (\ref{distributed estimator}). If Assumptions \ref{assumption1}--\ref{assumption3} hold, then $\forall i\in\mathcal N$,
\begin{align}\label{estimates}
    \Delta_i &\to -\frac{1}{N}\sum_{i=1}^N(A_ix_i-b_i(t)),\\
    y_i &\to -\left[\sum_{i=1}^N\rho_i\right]^{-1}\sum_{i=1}^N\phi_i,
\end{align}
and the settling time $T_1$ is upper bounded by $T_1^{\max}$.
\end{lemma}
\begin{proof}
See Appendix B.
\end{proof}

\begin{theorem}\label{theorem 1}
Consider problem (\ref{problem}) and the algorithm described in (\ref{distributed estimator})--(\ref{distributed controller}). Suppose that Assumptions 1--3 hold, and $\forall i\in\mathcal N$ that $\gamma_{3,\lambda}\geq\frac{N-1}{2}\Vert-\beta\Delta_i+\alpha_{i,\lambda}\Vert_2$ $\forall t\geq T_1^{\max}$ and $\gamma_{3,e}\geq\Vert A_i^\top\beta\Delta_i\Vert_2$ $\forall t\geq 2T_1^{\max}$. Then, the algorithm converges to the optimal trajectory within a fixed time. An upper bound on the convergence time $T_{sol}$ is
\begin{align}\label{T_sol}
    T_{sol}^{\max}&= 2T_1^{\max} + T_2^{\max}.
\end{align}
\end{theorem}

\begin{proof}
We employ a sequential Lyapunov analysis for this proof. Let $\bar y = \frac{1}{N}\sum_{i=1}^N y_i$ and $\bar\Delta = \frac{1}{N}\sum_{i=1}^N\Delta_i$. By Lemma 3, $y_i = \bar y = -\left[\sum_{i=1}^N\rho_i\right]^{-1}\sum_{i=1}^N\phi_i$ and $\Delta_i = \bar\Delta = -\frac{1}{N}\sum_{i=1}^N(A_ix_i-b_i(t))$ $\forall t\geq T_1^{\max}$ $\forall i\in\mathcal N$. In a similar manner, with $\gamma_{3,\lambda}\geq\frac{N-1}{2}\Vert-\beta\Delta_i+\alpha_{i,\lambda}\Vert_2$ for all $t\geq T_1^{\max}$, it can be shown that $\lambda_i = \bar\lambda$ $\forall t\geq 2T_1^{\max}$ $\forall i\in\mathcal N$.

Regarding $t\geq 2T_1^{\max}$, (\ref{distributed controller}) reads as
\begin{subequations}\label{reads as}
\begin{align}
    \dot x_i &= -F_{i,x} - f_{ixx}(x_i,t)^{-1}(A_i^\top\bar y + f_{ixt}(x_i,t)),\\
    \dot{\bar\lambda} &= -\sum_{i=1}^N\beta\bar\Delta + \bar y.
\end{align}
\end{subequations}
Fix any $i\in\mathcal N$. Consider the Lyapunov candidate $V = \frac{1}{2}\Vert e_i\Vert_2^2$. Using the chain rule, the time derivative of $V$ along (\ref{reads as}) can be obtained as
\begin{align}
    \dot V &= \langle e_i,f_{ixx}(x_i,t)\dot x_i +f_{ixt}(x_i,t) + A_i^\top\dot{\bar\lambda}\rangle \notag\\
    &= -\langle e_i,f_{ixx}(x_i,t)F_{i,x}+A_i^\top\beta\bar\Delta\rangle \notag\\
    &\leq^{(a)} -\gamma_1\Vert e_i\Vert^{2-\frac{p}{q}}_{2-\frac{p}{q}} - \gamma_2\Vert e_i\Vert^{2+\frac{p}{q}}_{2+\frac{p}{q}}\notag\\
    &\leq -2^{1-\frac{p}{2q}}\gamma_1V^{1-\frac{p}{2q}} - 2^{1+\frac{p}{2q}}n^{-\frac{p}{2q}}\gamma_2V^{1+\frac{p}{2q}},
\end{align}
where $\gamma_{3,e}\geq\Vert A_i^\top\beta\Delta_i\Vert_2$ $\forall t\geq 2T_1^{\max}$ has been used for deriving $\leq^{(a)}$. By Lemma 1, $e_i$ is fixed-time stable at the origin, and $e_i=\mathbf 0_n$ $\forall t\geq 2T_1^{\max} + T_2^{\max}$ $\forall i\in\mathcal N$.

Regarding $t\geq 2T_1^{\max} + T_2^{\max}$, we have
\begin{align}\label{convey KKT1}
    e_i &= f_{ix}(x_i,t) + A_i^\top\bar\lambda = \mathbf 0_n,\\
    \dot e_i &= -f_{ixx}(x_i,t)F_{i,x}-A_i^\top\beta\bar\Delta = \mathbf 0_n.
\end{align}
Substituting $e_i = \mathbf 0_n$ into (\ref{distributed controller}d) yields $F_{i,x} = \mathbf 0_n$, so that
\begin{align}\label{convey KKT2}
\bar\Delta = -(A_i^\top\beta)^{\dagger}\left(\dot e_i + f_{ixx}(x_i,t)F_{i,x}\right) = \mathbf 0_l.
\end{align}
As a result, we have $\dot x_i = \alpha_{i,x}$ and $\dot\lambda_i = \alpha_{i,\lambda}$, and (\ref{convey KKT1}) and (\ref{convey KKT2}) convey the KKT conditions for problem (\ref{problem}). In conjunction with Lemma 2, this implies that $x_i = x_i^\star$, $\lambda_i = \lambda^\star$, $\dot x_i = \dot x_i^\star$, and $\dot\lambda_i = \dot\lambda^\star$ $\forall t \geq T_{sol}^{\max}$ $\forall i\in\mathcal N$. The proof is complete.
\end{proof}

\begin{remark}\label{remark beta}
Regarding the feedback-feedforward algorithms proposed in this paper, only $\psi_i$, $\psi'_i$, $\Delta_i$, and $\lambda_i$ are exchanged among neighboring agents through the communication network. Neither the decision variables nor information about the local cost functions is shared. For the sake of (\ref{the following condition is consistently satisfied}), we employ (\ref{distributed controller}d) to drive $e_i$ to the origin within a fixed time, where $e_i=\mathbf 0_n$ corresponds to the stationarity condition. As indicated by (\ref{convey KKT1})--(\ref{convey KKT2}), the global equality constraints are satisfied and problem (\ref{problem}) is solved as soon as $e_i \to \mathbf 0_n$. In practice, digital implementation of the signum function induces chattering, which prevents $\dot e_i$ from vanishing exactly. One may observe $\sum_{i=1}^N(A_ix_i-b_i(t))$ converges to a neighborhood of the origin within a fixed time and then vanishes exponentially at rate $\beta$ \cite[Proposition 2]{8062794}, implying $\Vert\sum_{i=1}^N(A_ix_i(T_{sol})-b_i(T_{sol}))\Vert_2\propto 1/\beta$ from (\ref{convey KKT2}). Owing to (\ref{distributed estimator}f)--(\ref{distributed estimator}g), $\Delta_i\to -\frac{1}{N}\sum_{i=1}^N(A_ix_i-b_i(t))$ within a fixed time and it enters a small neighborhood of the origin at some $t$. This justifies the use of a relatively large $\beta$ in (\ref{distributed controller}e), while still satisfying the boundedness conditions in Theorem 1, which is important for mitigating the impact of chattering on convergence. The control parameters $\gamma_{3,(\cdot)}$ should be chosen sufficiently large to ensure convergence.
\end{remark}

\subsection{Feedback-Feedforward Algorithm with LFCs}
In this section, we extend the results of Section III.A and propose an initialization-free alternative to the barrier-based methods for handling LFCs, which are prevalent in practical applications but often omitted in related works. The goal is to find and track, in a distributed manner, the optimal trajectory $x^\star(t)$ defined by
\begin{align}\label{problem LFC}
\begin{split}
    x^\star(t) &= \underset{x(t)\in\mathbb R^{Nn}}{\mathrm{arg}\min} \sum_{i=1}^N f_i(x_i(t), t), \\
    \text{s.t. }& \sum_{i=1}^N A_i x_i(t) = \sum_{i=1}^N b_i(t), \\
    & g_{i,m}(x_i(t),t)\leq 0,\ \forall m\in\mathcal M, \ \forall i\in\mathcal N,
\end{split}
\end{align}
where $x_i(t)\in\mathbb R^{n}$, $b_i(t):\mathbb R_{+}\to\mathbb R^{l}$, $A_i\in\mathbb R^{l\times n}$, $g_{i,m}(x_i(t),t):\mathbb R^n\times\mathbb R_{+}\to\mathbb R$, and $\mathcal M = \{1,\dots,M\}$. This allows us to define the locally feasible set for agent $i$ as
\begin{align}\label{feasible set}
    X_i(t) = \{x_i\mid g_{i,m}(x_i,t) \leq 0,\ \forall m \in\mathcal M\},
\end{align}
whose interior and boundary are respectively given by
\begin{align}
    \operatorname{Int}(X_i(t)) &= \{X_i(t)\mid g_{i,m}(x_i,t) < 0,\ \forall m \in\mathcal M\},\\
    \partial X_i(t) &= X_i(t)\ \backslash \ \operatorname{Int}(X_i(t)).
\end{align}

\begin{assumption}\label{assumption4}
    The LFC $g_{i,m}(x_i,t)$ is uniformly convex with respect to $x_i$, twice continuously differentiable with respect to $x_i$, and continuously differentiable with respect to $t$. The Slater's condition holds at all times. 
\end{assumption}

For problem (\ref{problem LFC}), the presence of LFCs causes $x_i^\star$ to switch between $\operatorname{Int}(X_i(t))$ and $\partial X_i(t)$. In this context, $\dot x_i^\star$ is discontinuous at the switching instants and subject to differential inclusions. Nevertheless, we show that these time derivatives can still be comprehended through an equivalence between indicator functions and regularization terms.

\begin{lemma}\label{lemma optimal trajectory LFC}
Suppose Assumptions \ref{assumption1} and \ref{assumption4} hold. Let $x^\star$ and $\lambda^\star$ denote the optimal trajectory of primal and dual variables for problem (\ref{problem LFC}). Define the state-dependent switching signal as follows: $\sigma_{i,m}=1$ if $g_{i,m}(x_i,t)<0$ and $\sigma_{i,m}=0$ if $g_{i,m}(x_i,t)=0$. Let $\sigma_{i,\min} = \min_{m\in\mathcal M}\sigma_{i,m}$ and
\begin{subequations}\label{distributed estimator projection}
\begin{align}
    \rho_i &= A_iG_iA_i^\top,\\
    \phi_i &= A_iG_if_{ixt}(x_i,t)+ A_iH_i + b_{it}(t),
\end{align}  
where
\begin{align}
    G_i &= \sigma_{i,\min}f_{ixx}(x_i,t)^{-1},\\
    H_i &= \left[\sum_{m=1}^M (1-\sigma_{i,m})(g_{ixx,m} + g_{ix,m}g_{ix,m}^\top)\right]^{\dagger}\notag\\
    &\quad\times\sum_{m=1}^M (1-\sigma_{i,m})(g_{ixt,m} + g_{ix,m}g_{it,m}^\top).
\end{align}
\end{subequations}
Then, for almost all $t\geq 0$,
\begin{subequations}\label{extension projection}
\begin{align}
    \dot x_i^\star &= -G_i(A_i^\top\dot\lambda^\star+f_{ixt}(x_i^\star,t))-H_i,\\
    \dot\lambda^\star &= -\left[\sum_{i=1}^N\rho_i\right]^{\dagger}\sum_{i=1}^N\phi_i.
\end{align}
\end{subequations}
\end{lemma}

\vspace{2mm}

\begin{proof}
Associate each LFC with a regularization term $P(g_{i,m}(x_i,t),\eta):\mathbb R\times\mathbb R_{++}\to\mathbb R$, which is twice continuously differentiable and convex with respect to $g_{i,m}(x_i,t)$, and approaches the indicator function pointwise as $\eta\to+\infty$:
\begin{align}
    \lim_{\eta\to+\infty}P(g_{i,m}(x_i,t),\eta)=\mathbb I(g_{i,m}(x_i,t)),\notag
\end{align}
where $\mathbb I(g_{i,m}(x_i,t))=0$ for $g_{i,m}(x_i,t)<0$ and $I(g_{i,m}(x_i,t))=+\infty$ otherwise. Using this, we define the regularized local cost function:
\begin{align}
\widehat f_{i}(x_i,t,\eta) = f_{i}(x_i,t) + \sum_{m=1}^M P(g_{i,m}(x_i,t),\eta).
\end{align}
As $\eta\to+\infty$, the trajectory defined by $x^\star(t) = \mathrm{arg}\min\nolimits_{x(t)\in\mathbb R^{Nn}}\sum_{i=1}^N \widehat f_i(x_i(t),t,\eta) \text{ s.t. } \sum_{i=1}^NA_ix_i(t)=\sum_{i=1}^Nb_i(t)$ solves the original problem in (\ref{problem LFC}). Extending (\ref{analytical expressions}) with $\widehat f_i(x_i,t,\eta)$ leads to
\begin{subequations}\label{extension penalized}
\begin{align}
    \dot x_i^\star &= -\widehat f_{ixx}(x_i^\star,t,\eta)^{-1}(A_i^\top\dot\lambda^\star+\widehat f_{ixt}(x_i^\star,t,\eta)),\\
    \dot\lambda^\star &= -\left[\sum_{i=1}^NA_i\widehat f_{ixx}(x_i^\star,t,\eta)^{-1}A_i^\top\right]^{-1}\notag\\
    &\quad\times\left[\sum_{i=1}^NA_i\widehat f_{ixx}(x_i^\star,t,\eta)^{-1}\widehat f_{ixt}(x_i^\star,t,\eta)\right]\notag\\
    &\quad-\left[\sum_{i=1}^NA_i\widehat f_{ixx}(x_i^\star,t,\eta)^{-1}A_i^\top\right]^{-1}\sum_{i=1}^Nb_{it}(t).
\end{align}
\end{subequations}
By convexity, we obtain $\partial_{g_{i,m}} P(g_{i,m},\eta)\to\mathbb I(g_{i,m})$ and $\partial_{g_{i,m}}^2 P(g_{i,m},\eta)\to\mathbb I(g_{i,m})$ pointwise as $\eta\to+\infty$. Recalling (\ref{distributed estimator projection}c)--(\ref{distributed estimator projection}d), we arrive at the following relationship:
\begin{align}
    &\lim_{\eta\to+\infty}\widehat f_{ixx}(x_i,t,\eta)^{-1} = G_i,\notag\\
    &\lim_{\eta\to+\infty}\widehat f_{ixx}(x_i,t,\eta)^{-1}\widehat f_{ixt}(x_i,t,\eta) = G_if_{ixt}(x_i,t) + H_i.\notag
\end{align}
These convert (\ref{extension penalized}) into (\ref{extension projection}) and complete the proof.
\end{proof}

To this end, we propose a feedback-feedforward algorithm to solve problem (\ref{problem LFC}) and handle the LFCs. The distributed estimator retains the same structure as in (\ref{distributed estimator}), with $\rho_i$ and $\phi_i$ redefined as in (\ref{distributed estimator projection}a)--(\ref{distributed estimator projection}b). In terms of local feasibility, we define the discrepancy between $x_i$ and its projection \cite{9141512}:
\begin{align}
    \tilde x_i = x_i-\mathcal P_{X_i(t)}(x_i).
\end{align}
Then, the distributed controller incorporating projection-based feedback laws and feedforward laws that switch between different modes is as follows:
\begin{subequations}\label{distributed controller projection}
\begin{align}
    \dot x_i &= -F'_{i,x} + \alpha'_{i,x} - \gamma_1\tilde x_i^{1-\frac{p}{q}} - \gamma_2\tilde x_i^{1+\frac{p}{q}} - \gamma_{3,x}\operatorname{sign}(\tilde x_i)\notag\\
    &\quad  - \vert F'_{i,x}\vert\circ\operatorname{sign}(\tilde x_i),\\
    \dot\lambda_i &= - F_{i,\lambda} + \alpha_{i,\lambda} - \mathscr C_i(\lambda),\\
    e_i &= x_i - \mathcal P_{X_i(t)}(x_i - F_{i,x}),\\
    F'_{i,x} &= f_{ixx}(x_i,t)^{-1}(\gamma_1 e_i^{1-\frac{p}{q}}+\gamma_2 e_i^{1+\frac{p}{q}}+\gamma_{3,e}\operatorname{sign}(e_i)),\\
    F_{i,x} &= f_{ix}(x_i,t) + A_i^\top\lambda_i,\\
    F_{i,\lambda} &= \beta\Delta_i,\\
    \alpha_{i,x} &=
    -G_i(A_i^\top y_i+f_{ixt}(x_i,t)) - H_i,\\
    \alpha_{i,\lambda} &= y_i,
\end{align}
\end{subequations}
where $\alpha'_{i,x} = \alpha_{i,x}$ if $e_i = \mathbf 0_n$ and $\alpha'_{i,x}=\mathbf 0_n$ otherwise for a convergence guarantee. If $\tilde x_i=\mathbf 0_n$, (\ref{distributed controller projection}a) reduces to $\dot x_i = -F'_{i,x} + \alpha'_{i,x}$, and if $e_i=\mathbf 0_n$ as well, it further reduces to $\dot x_i = \alpha_{i,x}$. Moreover, $G_i$ and $H_i$ are given by (\ref{distributed estimator projection}), with an enhanced switching rule specified below:
\begin{align}\label{conservative switching rule}
\sigma_{i,m} =
\begin{cases}
    1 & \text{if } g_{i,m}(x_i-e_i,t) < 0,\\
    0 & \text{if } g_{i,m}(x_i-e_i,t) = 0.
\end{cases}
\end{align}

\begin{remark}\label{remark feasibility}
Note that (\ref{conservative switching rule}) acts as a more conservative switching rule without compromising optimality. It synchronizes with the projection operation and helps reduce the number of switches for converging to the optimal trajectory. According to (\ref{distributed controller projection}c), we have $x_i-e_i=\mathcal P_{X_i}(x_i-F_{i,x})\in X_i(t)$, which means $g_{i,m}(x_i-e_i,t) \leq 0$ at all times. Since $e_i\to \mathbf 0_n$ within a fixed time, it follows that $g_{i,m}(x_i-e_i,t)\to g_{i,m}(x_i,t)$ within a fixed time. As a result, the switching rule in (\ref{conservative switching rule}) achieves equivalence with the desired one, thereby preserving the optimality of the convergent solution. Furthermore, the switching of $\sigma_{i,m}$ is non-Zeno, \textit{i.e.}, there exists only a finite number of switching instants within any finite time interval. This property is detailed in Appendix C, where we establish the non-Zenoness of the switching rule and illustrate how it contributes to a reduction in the number of switches. Simulation results in Section \uppercase\expandafter{\romannumeral4}.A further validate the non-Zenoness.
\end{remark}

\begin{theorem}
Consider problem (\ref{problem LFC}) and the algorithm described in (\ref{distributed estimator}) and (\ref{distributed controller projection}).
Suppose that Assumptions \ref{assumption1}--\ref{assumption4} hold, and that $\gamma_{3,x}\geq\Vert\partial_t\mathcal P_{X_i(t)}(x_i)\Vert_2$ $\forall t\geq 0$ $\forall i\in\mathcal N$. For any initial value $x_i(0)\notin X_i(0)$, the trajectory of $x_i$ will enter $X_i(t)$ within a fixed time and remain in it thereafter, \textit{i.e.}, $x_i\in X_i(t)$ $\forall t\geq T_2^{\max}$. Moreover, the locally feasible set $X_i(t)$ is forward invariant if $x_i(0)\in X_i(0)$.
\end{theorem}
\begin{proof}
Fix any $i\in\mathcal N$. Define the distance-to-set: $d = \frac{1}{2}\Vert\tilde x_i\Vert_2^2$. On the one hand, $d = 0$ if $e_i=\mathbf 0_n$. On the other hand, for all $e_i\neq\mathbf 0_n$, applying the chain rule with the generalized gradient of projection gives
\begin{align}
    \dot d
    &= \langle\tilde x_i,\dot x_i-\partial_t\mathcal P_{X_i(t)}(x_i)\rangle\notag\\
    &\leq -\gamma_{1}\Vert\tilde x_i\Vert_{2-\frac{p}{q}}^{2-\frac{p}{q}} - \gamma_{2}\Vert\tilde x_i\Vert_{2+\frac{p}{q}}^{2+\frac{p}{q}}-\langle \tilde x_i,F'_{i,x}\rangle-\langle\vert\tilde x_i\vert,\vert F'_{i,x}\vert\rangle\notag\\
    &\quad -\gamma_{3,x}\Vert\tilde x_i\Vert_1 + \Vert\tilde x_i\Vert_2\Vert\partial_t\mathcal P_{X_i(t)}(x_i)\Vert_2\notag\\
    &\leq -2^{1-\frac{p}{2q}}\gamma_1d^{1-\frac{p}{2q}} - 2^{1+\frac{p}{2q}}n^{-\frac{p}{2q}}\gamma_2d^{1+\frac{p}{2q}},
\end{align}
provided that $\gamma_{3,x} \geq \Vert\partial_t\mathcal P_{X_i(t)}(x_i)\Vert_2$ $\forall t\geq 0$, as stated in Theorem 2. This establishes fixed-time stability of $d$ and that $x_i\in X_i(t)$ $\forall t\geq T_2^{\max}$ for any $x_i(0)\notin X_i(0)$. Moreover, suppose, for contradiction, that $x_i(t)$ leaves $X_i(t)$ at time $t_l$, which means $\lim_{t\to t_l^-}x_i(t)\in\partial X_i(t_l)$ and $\lim_{t\to t_l^+}x_i(t)\notin X_i(t_l+\epsilon)$. That is, $d(t_l)=0$ and $d(t_l+\epsilon)>0$ for $\epsilon\to 0^+$, a contradiction to $\dot d\leq 0$ which we have shown. As a result, once $x_i(t)$ enters $X_i(t)$, it will remain in it thereafter, making $X_i(t)$ forward invariant if $x_i(0)\in X_i(0)$. The proof is complete.
\end{proof}

According to (\ref{cone}), the condition $e_i = x_i - \mathcal P_{X_i(t)}(x_i - F_{i,x}) = \mathbf 0_n$ characterizes the stationarity of problem (\ref{problem LFC}), and can be equivalently written as
\begin{align}\label{stationarity condition}
\nu_i^\star + f_{ix}(x_i^\star,t) + A_i^\top\lambda^\star = \mathbf 0_n,\ \nu_i^\star\in\mathcal C_{X_i(t)}(x_i^\star),
\end{align}
where $\nu_i^\star$ is the feasible direction within the normal cone of $X_i(t)$ at $x_i^\star$. For almost all $t$, the time derivative of $\nu_i^\star$ is described by
\begin{align}
    \dot \nu_i^\star = -f_{ixx}(x_i^\star,t)\dot x_i^\star - f_{ixt}(x_i^\star,t) - A_i^\top\dot\lambda^\star.
\end{align}
Based on (\ref{distributed controller projection}), we define
\begin{align}\label{estimate of direction}
    \widehat{\dot \nu}_i = -f_{ixx}(x_i,t)\alpha'_{i,x} - f_{ixt}(x_i,t) - A_i^\top\alpha_{i,\lambda},
\end{align}
which serves as an estimate of $\dot \nu_i^\star$ since $\widehat{\dot \nu}_i=\dot \nu_i^\star$ for almost all $t$ if $x_i=x_i^\star$ and $\lambda_i=\lambda^\star$. Then, denote the switching instants of $\sigma_{i,m}$ $\forall m\in\mathcal M$ $\forall i\in\mathcal N$ as $0< t_1<t_2<\cdots<t_k<t_{k+1}<\cdots$. This construction enables us to state the following theorem.

\begin{theorem}
Consider problem (\ref{problem LFC}) and the algorithm described in (\ref{distributed estimator}) and (\ref{distributed controller projection}).
Suppose that Assumptions \ref{assumption1}--\ref{assumption4} hold, and $\forall i\in\mathcal N$ that $\gamma_{3,x}\geq\Vert\partial_t\mathcal P_{X_i(t)}(x_i)\Vert_2$ $\forall t\geq 0$, $\gamma_{3,\lambda}\geq\frac{N-1}{2}\Vert -\beta\Delta_i+\alpha_{i,\lambda}\Vert_2$ $\forall t\geq T_1^{\max}$, and $\gamma_{3,e}\geq\Vert \widehat{\dot \nu}_i+A_i^\top\beta\Delta_i\Vert_2$ $\forall t\geq \max\{2T_1^{\max},T_2^{\max}\}$. Then, the following statements are true for any $x_i(0)\in\mathbb R^n$ and $\lambda_i(0)\in\mathbb R^l$:
\begin{enumerate}
    \item For all $i,j\in\mathcal N$, $\Delta_i=\Delta_j$ $\forall t\geq T_1^{\max}$ and $\lambda_i=\lambda_j$ $\forall t\geq 2T_1^{\max}$. For all $i\in\mathcal N$, $x_i\in X_i(t)$ $\forall t\geq T_2^{\max}$ and $e_i=\mathbf 0_n$ $\forall t\geq \max\{2T_1^{\max},T_2^{\max}\} + T_2^{\max}$.
    \item The algorithm converges to the optimal trajectory within a fixed time for consecutive switching instants satisfying $t_{k+1}-t_{k}\geq T_{sol}$, where $k\in\mathbb Z_{+}$ and $t_0 = 0$. An upper bound on the convergence time $T_{sol}$ is $T_{sol}^{\max} = \max\{2T_1^{\max} + T_2^{\max},2T_2^{\max},t_k + T_1^{\max}\} - t_k$;
    \item The algorithm is globally asymptotically stable at the optimal trajectory if the switching sequence is finite.
\end{enumerate}
\end{theorem}

\begin{proof}
A sequential Lyapunov analysis is performed for this proof, beginning by establishing fixed-time consensus of $\Delta_i$ and $\lambda_i$. Repeating the steps in Section \uppercase\expandafter{\romannumeral3}.A, we have $\Delta_i=\bar\Delta= -\frac{1}{N}\sum_{i=1}^N(A_ix_i-b_i(t))$ $\forall t\geq T_1^{\max}$ $\forall i\in\mathcal N$, $\lambda_i=\bar\lambda$ $\forall t\geq 2T_1^{\max}$ $\forall i\in\mathcal N$, and $x_i\in X_i(t)$ $\forall t\geq T_2^{\max}$. Notably, finding $x_i^\star$ for $\mathcal P_{X_i(t)}(x_i - F_{i,x})= x_i$ is equivalent to finding $x_i^\star$ for the variational inequality: $\langle F_{i,x},x_i-x_i^\star\rangle \geq 0$ $\forall x_i\in X_i(t)$. This allows for convergence analysis of $e_i$ leveraging variational inequality theory \cite{fukushima1992equivalent}.

Fix any $i\in\mathcal N$. Regarding $t\geq \max\{2T_1^{\max},T_2^{\max}\}$, consider the Lyapunov function
\begin{align}\label{V lower bound}
    V &= \langle F_{i,x}-e_i,e_i\rangle  + \frac{1}{2}\Vert e_i\Vert_2^2 \geq \frac{1}{2}\Vert e_i\Vert_2^2.
\end{align}
By \cite[Theorem 3.2]{fukushima1992equivalent}, $\langle F_{i,x}-e_i,e_i\rangle \geq 0$ holds because $x_i\in X_i(t)$, and $V = -\min_{x'_i\in X_i(t)}V'$ with $V' = -\langle F_{i,x},x_i-x'_i\rangle  + \frac{1}{2}\Vert x_i-x'_i\Vert_2^2$. Here the minimum is uniquely attained at $x'_i=\mathcal{P}_{X_i(t)}(x_i-F_{i,x})$, which yields $x_i-x'_i = e_i$ and $\partial_{x_i'}V'=0$. Using the chain rule, we have
\begin{align}
\dot V &= -\langle\partial_{x_i}V',\dot x_i\rangle - \langle\partial_{\bar\lambda}V',\dot{\bar\lambda}\rangle - \partial_{t}V'\\
&= \underbrace{\langle e_i,\partial_{x_i} F_{i,x}\dot x_i + \partial_t F_{i,x} + \partial_{\bar\lambda} F_{i,x}\dot{\bar\lambda}\rangle}_{\dot V_1} + \underbrace{\langle F_{i,x}-e_i,\dot x_i\rangle }_{\dot V_2}.\notag
\end{align}
The first term can be obtained as
\begin{align}\label{dot Vi1}
    \dot V_1 &= -\langle e_i,\gamma_1 e_i^{1-\frac{p}{q}}+\gamma_2 e_i^{1+\frac{p}{q}}+\gamma_{3,e}\operatorname{sign}(e_i)\rangle \notag\\
    &\quad - \langle e_i,\widehat{\dot \nu}_i+A_i^\top\beta\bar\Delta\rangle \notag\\
    &\leq^{(b)} -\gamma_1\Vert e_i\Vert^{2-\frac{p}{q}}_{2-\frac{p}{q}} - \gamma_2\Vert e_i\Vert^{2+\frac{p}{q}}_{2+\frac{p}{q}},
\end{align}
where $\gamma_{3,e}\geq\Vert\widehat{\dot \nu}_i + A_i^\top\beta\Delta_i\Vert_2$ $\forall t\geq \max\{2T_1^{\max},T_2^{\max}\}$ $\forall i\in\mathcal N$ has been used for deriving $\leq^{(b)}$.

The normal cone of $X_i(t)$ at $P_{X_i(t)}(x_i-F_{i,x})$ is $\mathcal C_{X_i(t)}(\mathcal P_{X_i(t)}(x_i-F_{i,x})) = \{\nu_i\in\mathbb R^n \mid \langle \nu_i, x'_i - \mathcal P_{X_i(t)}(x_i-F_{i,x})\rangle \leq 0,\ \forall x'_i\in X_i(t)\}$.
Clearly, $-(F_{i,x}-e_i)\in\mathcal C_{X_i(t)}(\mathcal P_{X_i(t)}(x_i-F_{i,x}))$ since $\langle F_{i,x}-e_i,e_i\rangle\geq 0$ and $x_i-\mathcal P_{X_i(t)}(x_i-F_{i,x}) = e_i$. Given $x_i\in X_i(t)$, there always exists a sufficiently small $h>0$ such that $x'_i=x_i-h\operatorname{sign}(e_i)\in X_i(t)$. Recalling the definition of normal cone yields $\langle F_{i,x}-e_i,x_i-x_i+h\operatorname{sign}(e_i)\rangle \geq 0$ and hence $\langle F_{i,x}-e_i,\operatorname{sign}(e_i)\rangle \geq 0$. Also, it geometrically implies $\langle F_{i,x}-e_i,e_i^{1\pm\frac{p}{q}}\rangle\geq 0$. Therefore,
\begin{align}\label{dot Vi2}
    \dot V_2 &= -\langle F_{i,x}-e_i,\gamma_1 e_i^{1-\frac{p}{q}}+\gamma_2 e_i^{1+\frac{p}{q}}+\gamma_{3,e}\operatorname{sign}(e_i)\rangle \notag\\
    &\leq 0,\ \forall e_i\neq\mathbf 0_n.
\end{align}

Combining (\ref{V lower bound})--(\ref{dot Vi2}), for all $e_i\neq\mathbf 0_n$, again we have
\begin{align}
    \dot V
    &\leq -2^{1-\frac{p}{2q}}\gamma_1V^{1-\frac{p}{2q}} - 2^{1+\frac{p}{2q}}n^{-\frac{p}{2q}}\gamma_2V^{1+\frac{p}{2q}}.
\end{align} 
By Lemma 1 and $V\geq\frac{1}{2}\Vert e_i\Vert_2^2$, we know $e_i$ is fixed-time stable at the origin:
\begin{align}\notag
    e_i =  x_i - \mathcal P_{X_i}(x_i - F_{i,x}) = \mathbf 0_n,
\end{align}
for all $t\geq \max\{2T_1^{\max},T_2^{\max}\} + T_2^{\max}$ and $i\in\mathcal N$. This implies a direction $\nu_i\in\mathcal C_{X_i(t)}(x_i)$ such that $\nu_i + f_{ix}(x_i,t) + A_i^\top\bar\lambda = \mathbf 0_n$. Taking the time derivative, we obtain
\begin{align}\label{a solution to}
    -A_i^\top\beta\bar\Delta + \dot \nu_i-\widehat{\dot \nu}_i = \mathbf 0_n.
\end{align}   
Regarding $t\geq \max\{2T_1^{\max},T_2^{\max}\}+T_2^{\max}$, (\ref{distributed controller projection}a)--(\ref{distributed controller projection}b) can be reduced to $\dot x_i \in \alpha_{i,x}$ and $\dot\lambda_i \in - F_{i,\lambda} + \alpha_{i,\lambda}$, which are locally Lipschitz between consecutive switching instants. A solution to (\ref{a solution to}) exists at the argument of $\nu_i + f_{ix}(x_i,t) + A_i^\top\bar\lambda = \mathbf 0_n$, $-A_i^\top\beta\bar\Delta = \mathbf 0_n$, and $\dot \nu_i-\widehat{\dot \nu}_i = \mathbf 0_n$. While $\nu_i + f_{ix}(x_i,t) + A_i^\top\bar\lambda = \mathbf 0_n$ and $-A_i^\top\beta\bar\Delta = \mathbf 0_n$ convey the KKT conditions for problem (\ref{problem LFC}), we have $\dot \nu_i \to \dot \nu_i^\star$ and $\widehat{\dot \nu}_i \to \dot \nu_i^\star$ as $x_i \to x_i^\star$ and $\bar\lambda \to \lambda^\star$ if and only if $y_i \to \bar y$. As per Picard-Lindelöf theorem, the convergent solution constitutes the unique solution to (\ref{a solution to}) and the optimal solution to problem (\ref{problem LFC}). Similar to the proof of Lemma \ref{lemma distributed estimator}, we can show $y_i = -\left[\sum_{i=1}^N\rho_i\right]^{-1}\sum_{i=1}^N\phi_i$ for any two consecutive switching instants satisfying $t_{k+1}-t_{k}\geq T_{sol}$. Therefore, the algorithm converges within a fixed time, and the convergence time $T_{sol}$ is upper bounded by $T_{sol}^{\max} = \max\{2T_1^{\max} + T_2^{\max},2T_2^{\max},t_k + T_1^{\max}\} - t_k$. Overall, the algorithm is globally asymptotically stable at the optimal trajectory if there are finitely many switching instants. The concludes the proof.
\end{proof}

\begin{remark}
The proposed algorithm can be extended to handle global affine inequality constraints as well. This is because such constraints can be reformulated as a combination of global affine equality constraints and LFCs, as studied in Section \uppercase\expandafter{\romannumeral3}.B. To illustrate, consider the problem of finding and tracking $x^\star(t) = \mathrm{arg}\min\nolimits_{x(t)\in\mathbb R^{Nn}}\sum_{i=1}^N f_i(x_i(t),t)$ s.t. $\sum_{i=1}^N(A_ix_i(t)-b_i(t)) \preceq \mathbf 0_l$. By introducing a slack variable $\bar s(t)\succeq \mathbf 0_l$, the global affine inequality constraints can be replaced by $\sum_{i=1}^N(A_ix_i-b_i(t))+\bar s = \mathbf 0_l$. Here, $\bar s$ can be distributed as $s_i\succeq \mathbf 0_l$ with $s_i=s_j$ for all $i,j\in\mathcal N$, which is equivalent to projecting $s_i$ onto the right-half plane and maintaining their consensus. To this end, the projection-based method in Section \uppercase\expandafter{\romannumeral3}.B can be adopted. Handling global nonlinear inequality constraints may require extensive modifications, which will be part of our future work.
\end{remark}

\section{Case Studies}
This section presents simulation results obtained in MATLAB/SIMULINK with a fixed step size of $0.1$ ms, covering two numerical examples and a power system application.

\subsection{Numerical Examples}
A multi-agent system with $N=6$ agents is considered for the numerical examples. The communication network is connected, undirected, and fixed for all $t\geq 0$, as visualized in Fig. 1. The second smallest eigenvalue of the Laplacian matrix $\mathcal L$ is $\eta_2(\mathcal L) = 3$. All the local cost functions and activity functions are time-varying. For $i=1,\dots,6$, we set $A_i = 1$, $f_i(x_i,t) = (1+0.1i)x_i^2 + 0.2\sin(0.1it)x_i^2$, and $b_i(t) = 10i + 5\sin(0.1it) + 0.1it$. The initial values are $x_i(0)=\lambda_i(0) = 0$ for all $i\in\mathcal N$. As discussed in Remark \ref{remark singularity}, we additionally set $y_i$ to zero whenever $\psi_i<0.1$. The control parameters are selected as $p=2$, $q=3$, $\beta = 50$, $\gamma_1=\gamma_2 = 10$, $\gamma_{3,\psi}=\gamma_{3,\psi'} = 1$, and $\gamma_{3,\Delta}=\gamma_{3,\lambda}=\gamma_{3,e} = 100$. From (\ref{T1})--(\ref{T2}), we obtain $T_1^{\max} = 0.1427$ s and $T_2^{\max} = 0.2356$ s.
\begin{figure}[htbp]
    \centering
    \includegraphics[width=0.5\linewidth]{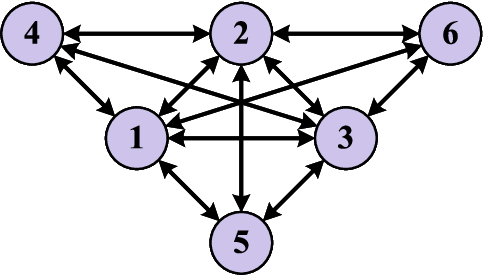}
    \caption{Topology of the communication network.}
\end{figure}

\begin{figure}[t]
    \centering
    \begin{subfigure}{0.4\textwidth}
        \centering
        \includegraphics[width=\textwidth]{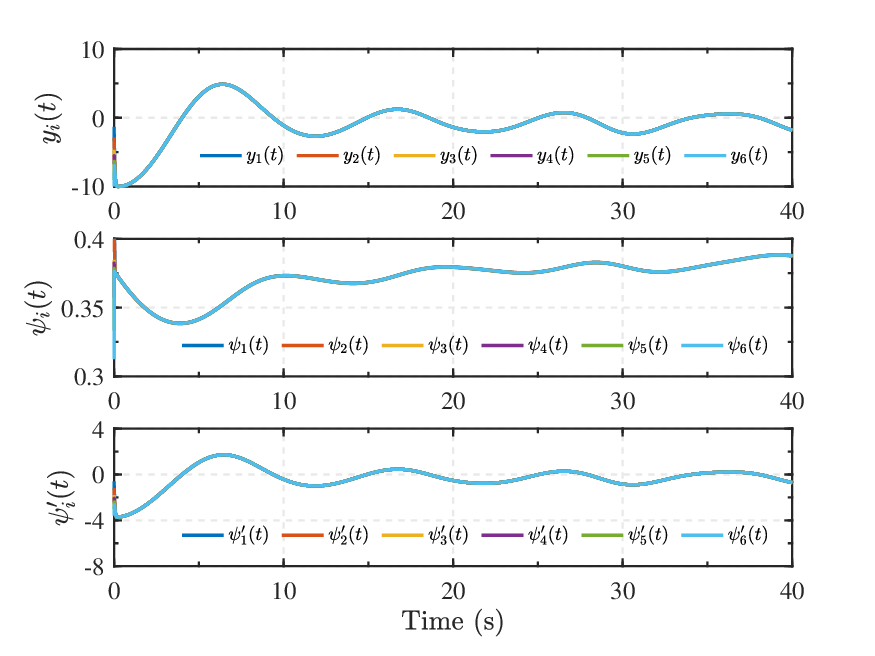}
        \caption{}
    \end{subfigure}
    
    \begin{subfigure}{0.24\textwidth}
        \centering
        \includegraphics[width=\textwidth]{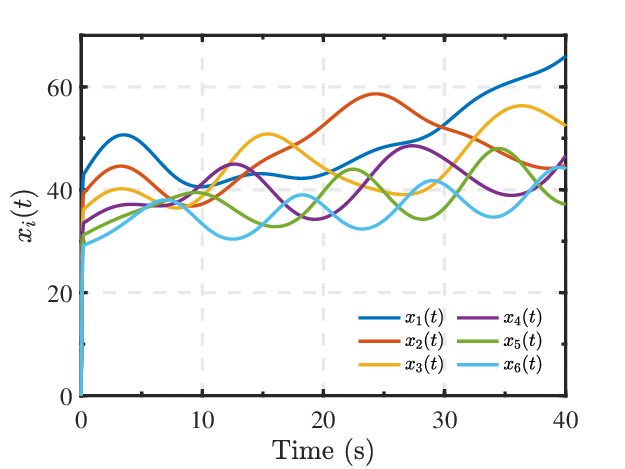}
        \caption{}
    \end{subfigure}
    \begin{subfigure}{0.24\textwidth}
        \centering
        \includegraphics[width=\textwidth]{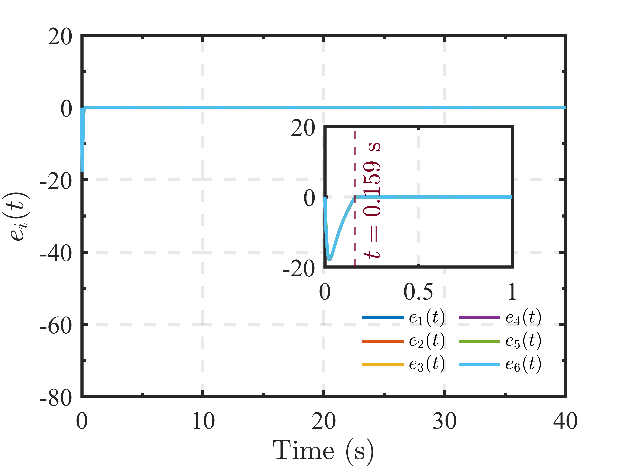}
        \caption{}
        \label{subfig:Proj_e}
    \end{subfigure}
 
    \begin{subfigure}{0.24\textwidth}
        \centering
        \includegraphics[width=\textwidth]{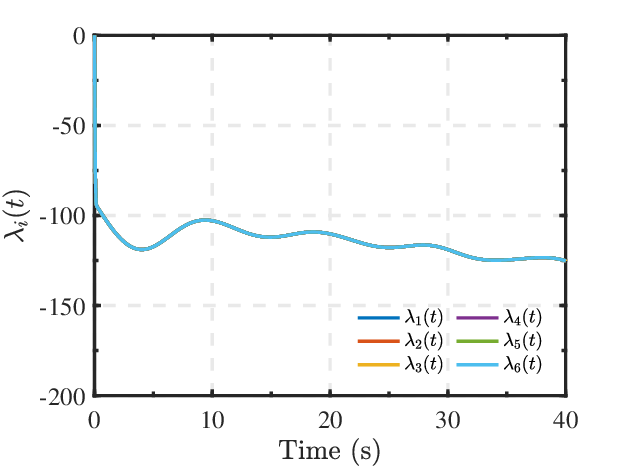}
        \caption{}
    \end{subfigure}
    \begin{subfigure}{0.24\textwidth}
        \centering
        \includegraphics[width=\textwidth]{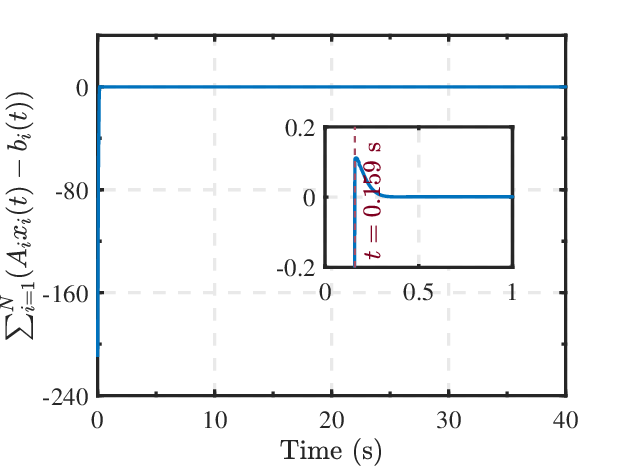}
        \caption{}
    \end{subfigure}
    \caption{Evolution of (a) local estimates; (b) decision variable; (c) error with respect to the stationarity condition; (d) dual variable; and (e) violation of the global equality constraint for Numerical Example 1.}
\end{figure}

\begin{figure}[t]
    \centering
    \begin{subfigure}{0.4\textwidth}
        \centering
        \includegraphics[width=\textwidth]{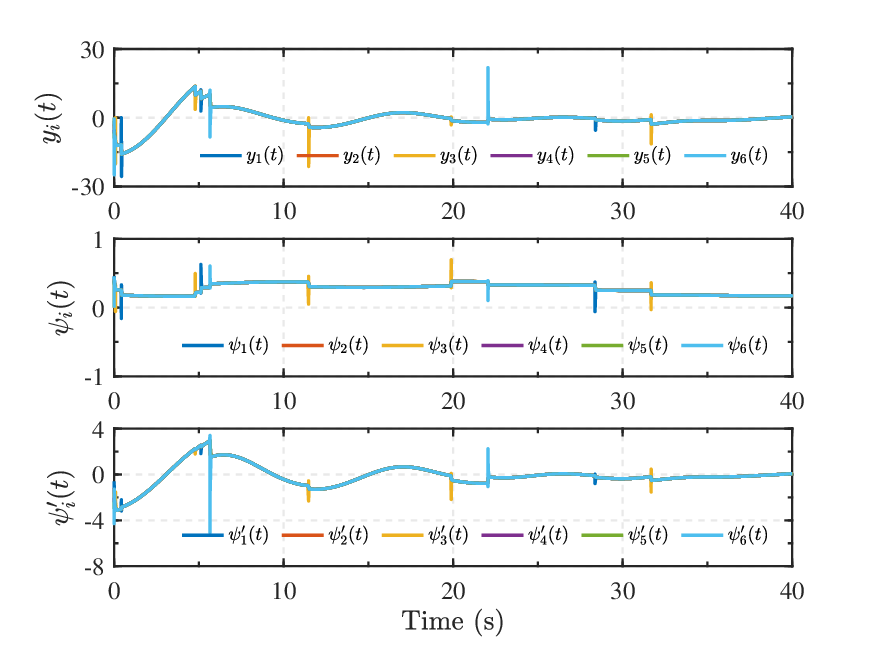}
        \caption{}
    \end{subfigure}
    
    \begin{subfigure}{0.24\textwidth}
        \centering
        \includegraphics[width=\textwidth]{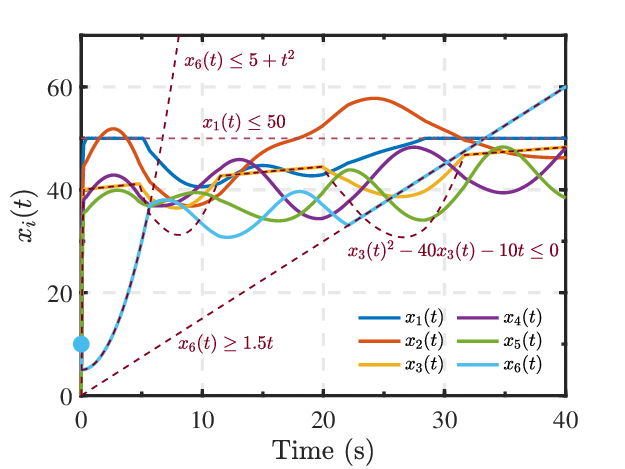}
        \caption{}
    \end{subfigure}
    \begin{subfigure}{0.24\textwidth}
        \centering
        \includegraphics[width=\textwidth]{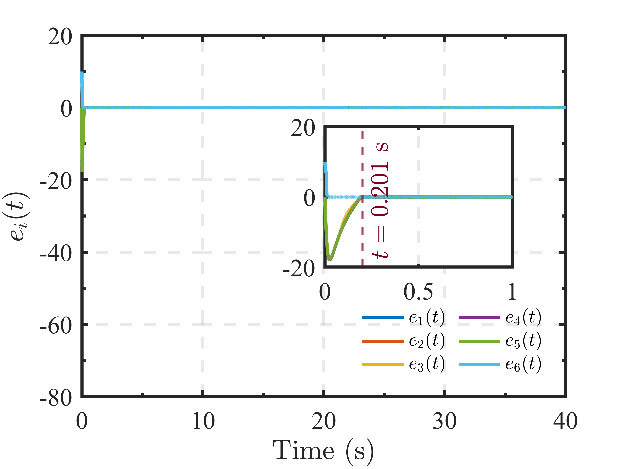}
        \caption{}
        \label{subfig:Proj_e}
    \end{subfigure}
 
    \begin{subfigure}{0.24\textwidth}
        \centering
        \includegraphics[width=\textwidth]{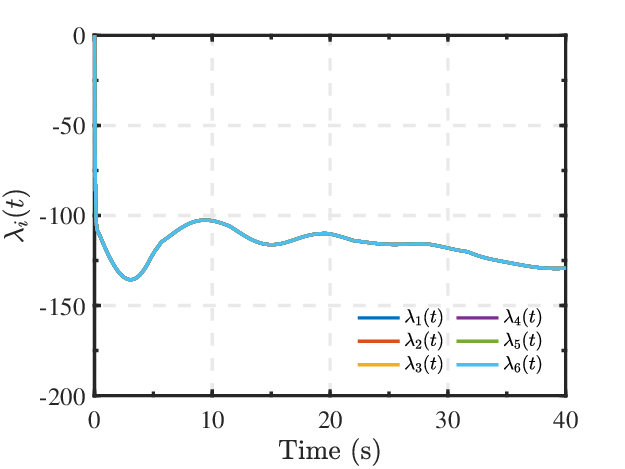}
        \caption{}
    \end{subfigure}
    \begin{subfigure}{0.24\textwidth}
        \centering
        \includegraphics[width=\textwidth]{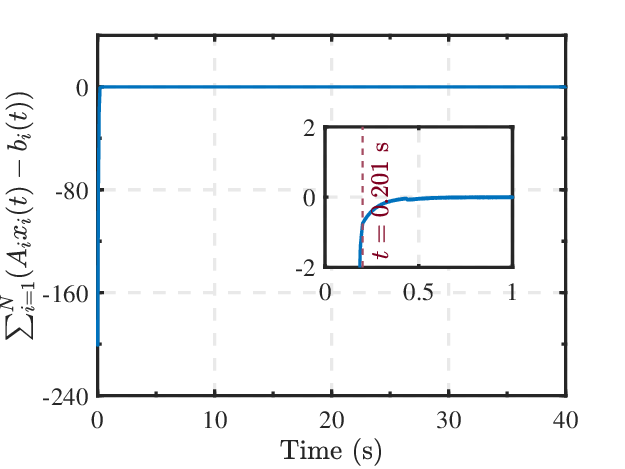}
        \caption{}
    \end{subfigure}
    \caption{Evolution of (a) local estimates; (b) decision variable; (c) error with respect to the stationarity condition; (d) dual variable; and (e) violation of the global equality constraint for Numerical Example 2.}
    \label{fig:Proj}
\end{figure}

We begin with Numerical Example 1 where LFCs are absent. Problem (\ref{problem}) is solved using the algorithm proposed in Section \uppercase\expandafter{\romannumeral3}.A, and the simulation results are shown in Fig. 2. The local estimates $\psi_i$ and $\psi'_i$ for all $i\in\mathcal N$ reach consensus at $t=0.005$ s and $t=0.045$ s, respectively. As a result, $y_i = y_j$ holds for all $t\geq 0.045$ s and $i,j\in\mathcal N$, providing a valid estimate of $-[\sum_{i=1}^N\rho_i]^{-1}\sum_{i=1}^N\phi_i$ as required. Similarly, $\lambda_i$ reaches consensus across all agents at $t = 0.010$ s. Both $x_i$ and $\lambda_i$ exhibit significant temporal variations due to the time-varying nature of $f_i(x_i,t)$ and $b_i(t)$. Moreover, $e_i$ converges to zero at $t = 0.159$ s; theoretically, $\sum_{i=1}^N(A_ix_i-b_i(t)) = 0$ should be met simultaneously, as implied by fixed-time stability. However, due to the impact of chattering, $\sum_{i=1}^N(A_ix_i-b_i(t))$ instead converges to a small neighborhood around the origin, with a minor residual of $0.11$ observed at $t=0.159$ s, which then vanishes exponentially. Since $e_i=0$ and $\sum_{i=1}^N(A_ix_i-b_i(t))=0$ convey the KKT conditions, problem (\ref{problem}) is considered solved within $T_{sol}=0.159$ s, where $T_{sol}^{\max}=0.521$ s.

We proceed with Numerical Example 2, where LFCs are present. Problem (\ref{problem LFC}) is solved the algorithm proposed in Section \uppercase\expandafter{\romannumeral3}.B, and the simulation results are shown in Figs. 3--4. The problem setup and control parameters are identical to those in Numerical Example 1, except for the inclusion of $\gamma_{3,x}=10$ and the following LFCs: $x_1(t)\leq 50$, $x_3(t)^2 - 40x_3(t) - 10t\leq 0$, and $1.5t\leq x_6(t)\leq 5 + t^2$. To demonstrate that the algorithm is initialization-free, the initial values are set as $x(0) = [0;0;0;0;0;10]$ and $\lambda(0) = [0;0;0;0;0;0]$. Clearly, $x_6(0)\notin X_6(0)$. When implementing the switching rule in (\ref{conservative switching rule}), computer precision must be taken into account, whether in Matlab or other platforms. To avoid inconsistencies and numerical difficulties, for example, we set $\sigma_{i,m}=1$ if $g_{i,m}(x_i-e_i)< -10^{-15}$ and $\sigma_{i,m}=0$ otherwise. The simulation results are shown in Figs. 3--4. A total of $20$ switches are recorded over the simulated time span. The local estimates are subject to the state-dependent switching signal; consequently, $\psi_i$ and $\psi'_i$ exhibit discontinuities at the switching instants and have to re-establish consensus after each switch. In general, $y_i=-[\sum_{i=1}^N\rho_i]^{-1}\sum_{i=1}^N\phi_i$ can be recovered within approximately $0.05$ s (the upper bound is $T_1^{\max} = 0.1427$ s). Regarding $i=1,\dots,5$, $x_i\in X_i(t)$ for all $t\geq 0$. Starting from an infeasible initial value, $x_6$ enters the locally feasible set at $t=0.034$ s (the upper bound is $T_2^{\max} = 0.2356$ s) and remains within it thereafter, without further violations of the LFCs---demonstrating the initialization-free capability of our projection-based method. Furthermore, $\lambda_i$ reaches consensus across all agents at $t=0.010$ s, and $e_i$ converges to zero at $t=0.201$ s. Similar to Numerical Example 1, $\sum_{i=1}^N(A_ix_i-b_i(t))$ exhibits a minor residual of $-0.74$ at $t=0.201$ s before vanishing exponentially. The switching rule is non-Zeno, as only finitely many switches occur within a finite time interval. Moreover, since $\sum_{i=1}^N(A_ix_i-b_i(t))$ is only minimally disturbed and rapidly restored to zero, the effects of switching on the satisfaction of the global equality constraint are negligible. Given the switching instants $t_{3}=0.082$ s and $t_{4}=0.435$ s, problem (\ref{problem LFC}) is considered solved at $t=0.201$ s, with $T_{sol} = 0.119$ s and $T_{sol}^{\max} = 0.439$ s.

\begin{figure}[t]
    \centering
    \includegraphics[width=0.4\textwidth]{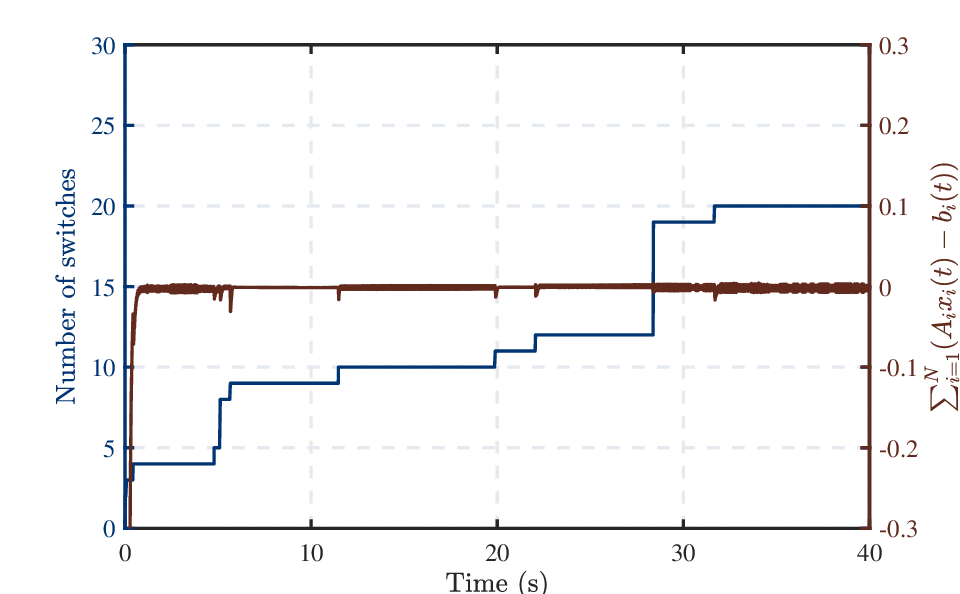}
    \caption{Non-Zenoness of switching and its negligible effects on global equality constraint satisfaction.}
\end{figure}

\subsection{Power System Application}
In what follows, we demonstrate the effectiveness and scalability of the projection-based method via a power system application. Different from the numerical examples, this distributed TVRA problem varies with frequency and thus implicitly with time.

\begin{figure}[b]
	\centering
        \includegraphics[width=0.4\textwidth]{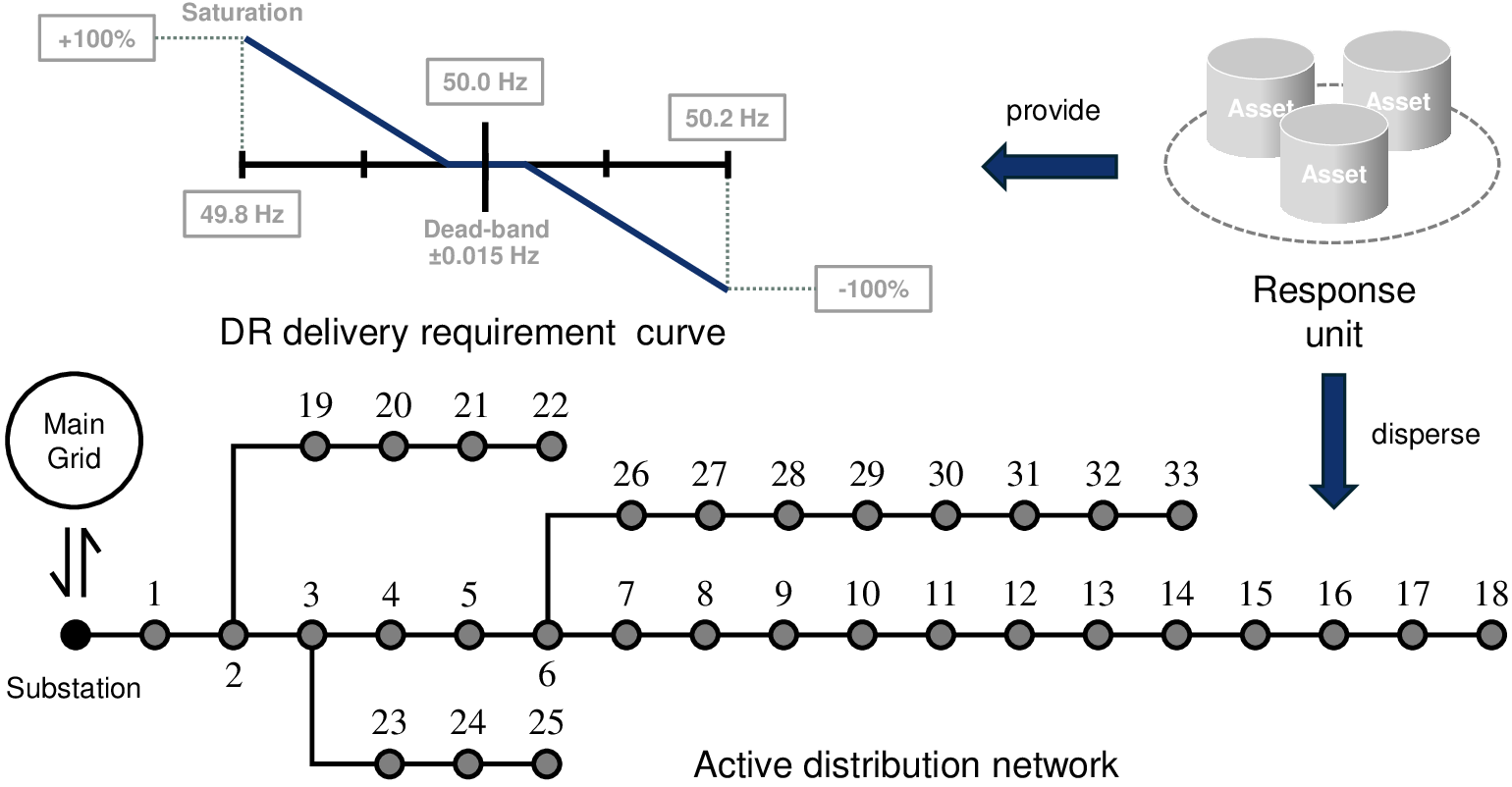}
	\caption{Schematic overview of distributed TVRA for DR.}
	\label{case}
\end{figure}
\begin{figure}[t]
    \centering
    \includegraphics[width=0.4\textwidth]{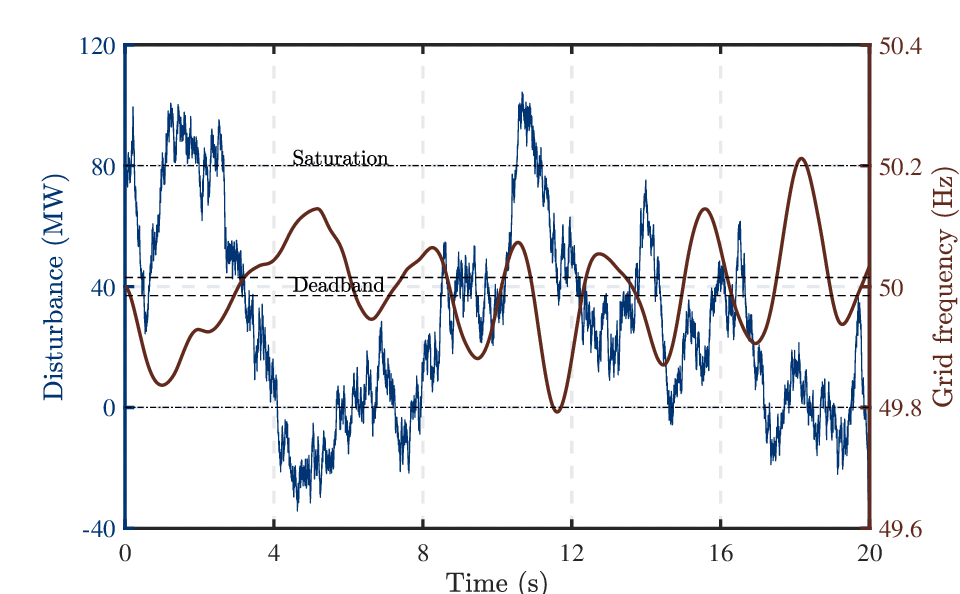}
    \caption{Disturbance and evolution of grid frequency.}
\end{figure}
\begin{figure}[h]
\centering
    \begin{subfigure}{0.4\textwidth}
    \includegraphics[width=\textwidth]{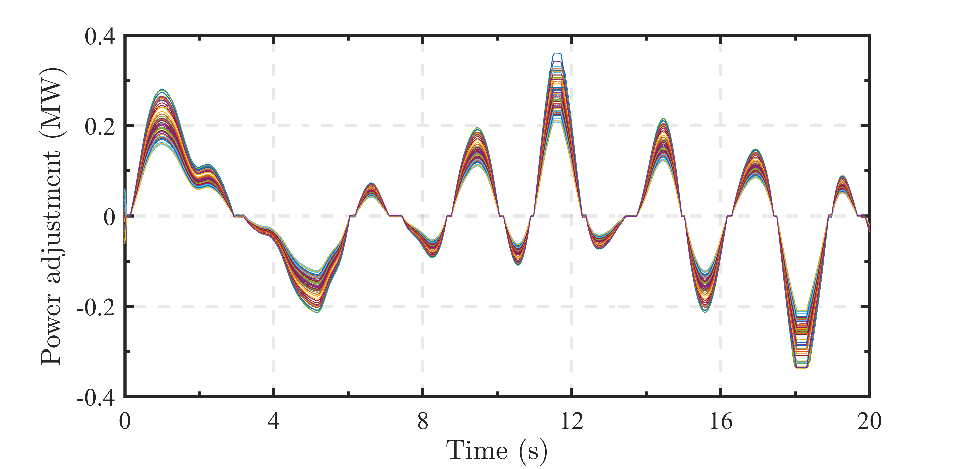}
    \caption{}
    \end{subfigure}

    \begin{subfigure}{0.4\textwidth}
    \includegraphics[width=\textwidth]{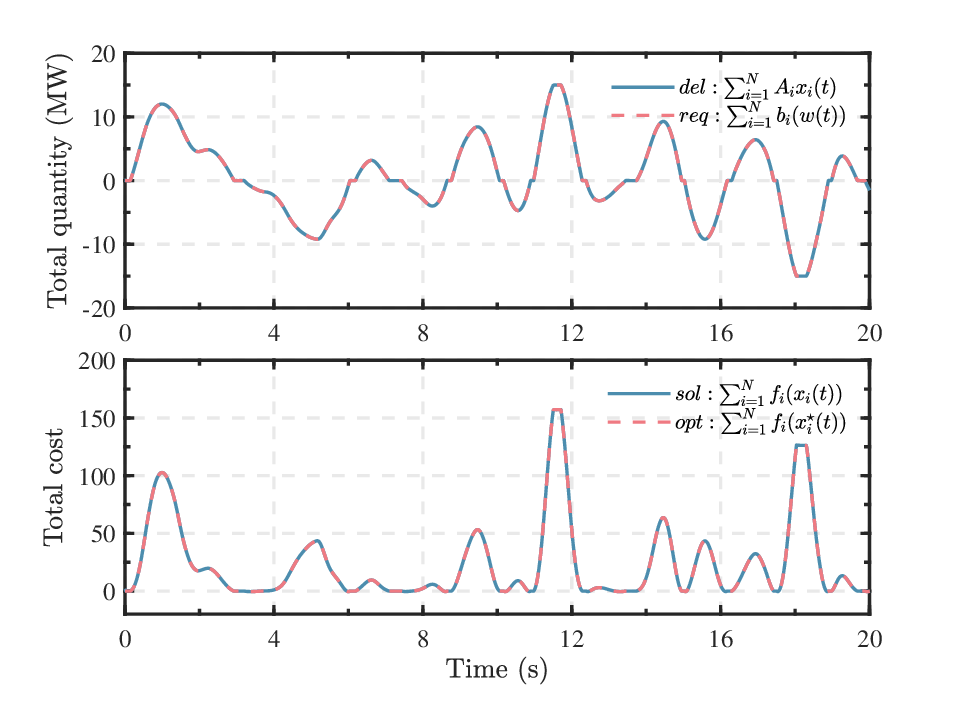}
    \caption{}
    \end{subfigure}
    \caption{Evolution of (a) power adjustment of each asset; and (b) delivered quantity and cost of the response unit.}
\end{figure}

Variations in instantaneous grid frequency serve as an indicator of power imbalance. The growing penetration of renewable energy sources (RESs) poses significant challenges to frequency stability in modern power systems due to their stochastic nature \cite{9796617}. For stability and power quality, the grid frequency must be maintained within statutory limits, and to this end, the UK’s National Energy System Operator (NESO) has recently introduced a suite of dynamic services aimed at improving fast frequency response. These services are designed with different response envelopes to collectively maintain the grid frequency within 50$\pm$0.5 Hz \cite{NGESO_guide}. 

Among these services, Dynamic Regulation (DR) is a pre-fault service for correcting continuous but small deviations in grid frequency. Launched in 2022, DR allows eligible assets to aggregate into a response unit that provides fast frequency response to the main grid. As depicted in Fig. 5, the quantity required by DR is determined by the real-time frequency deviation and the contracted quantity of the response unit \cite{NGESO_guide}:
\begin{align}
b(w(t)) =
\begin{cases}
    0 & \text{case 1},\\
    -5.405(w(t)+0.015)q_c & \text{case 2},\\
    5.405(w(t)-0.015)q_c & \text{case 3},\\
    q_c & \text{case 4},\\
\end{cases}
\end{align}
where $b(w(t)):\mathbb R\to\mathbb R$ is the required quantity in MW, $w(t)\in\mathbb R$ the grid frequency deviation in Hz, and $q_c\in\mathbb Z_{++}$ the contracted quantity in MW. Case 1 is when $-0.015< w(t)<0.015$; case 2 is when $-0.2< w(t)\leq-0.015$; case 3 is when $0.015\leq w(t)<0.2$; and case 4 is when $w(t)\leq-0.2$ or $w(t)\geq0.2$. These conditions give rise to a global equality constraint that must be satisfied collectively by all assets in the response unit:
\begin{align}
    \sum_{i=1}^NA_ix_i=b(w(t)),
\end{align}
where $A_i\in\mathbb R$ represents the energy efficiency factor accounting for distribution losses, and $x_i(t)\in\mathbb R$ is the quantity delivered by adjusting power from the operational baseline. The main grid is a multi-machine system \cite{shi2018analytical}, governed by the generalized swing equation \cite{kundur}:
\begin{align}
    \frac{2H}{w_s}\dot w(t) = -Dw(t) + \Delta P_m(t) - \Delta P_e(t),
\end{align}
where $H\in\mathbb R_{++}$ and $D\in\mathbb R_{++}$ are the equivalent inertia and damping constants, respectively, and $w_s = 50$ Hz is the nominal frequency. With the pre-disturbance state taken as the steady state, $\Delta P_m(t)\in\mathbb R$ and $\Delta P_e(t)\in\mathbb R$ denote the deviations of mechanical power and electrical power (in p.u.) from the steady-state values. It is noted that $\Delta P_e(t)$ includes components from load, RESs, and assets. Each machine is equipped with turbine governor control and secondary frequency control. Detailed system modeling and parameters can be found in \cite{shi2018analytical}.

Based on the service specifications of DR, problem (\ref{problem LFC}) is modified to optimally provide DR as follows: $x^\star(t) = \mathrm{arg}\min\nolimits_{x(t)\in\mathbb R^N}\sum_{i=1}^N f_i(x_i(t)) \text{ s.t. } \sum_{i=1}^NA_ix_i(t)=b(w(t)) \text{ and } x_i^{\min}\leq x_i(t)\leq x_i^{\max}\ \forall i\in\mathcal N$, where $f_i(x_i(t))$ is a quadratic cost function penalizing power adjustment from the operational baseline, and $x_i^{\min},x_i^{\max}\in\mathbb R$ are limits on the power adjustment. The algorithm proposed in Section \uppercase\expandafter{\romannumeral3}.B is employed for solving this problem, with the following modifications: $b_i(t) = \frac{1}{N}b(w(t))$ and $b_{it}(t) = \frac{1}{N}\nabla b(w(t))\dot w(t)$. With a contracted quantity of $q_c=15$ MW, $N=60$ assets are dispersed across $33$ buses. The communication network is configured such that $a_{ij}=1$ for all $\vert i-j\vert=1,2,3$ and $a_{ij}=0$ otherwise. The control parameters remain unchanged from Section \uppercase\expandafter{\romannumeral4}.A. As shown in Fig. 6, stochastic fluctuations in load and RES power perform the source of disturbances. These fluctuations lead to variations in the grid frequency, which, in turn, cause $x_i$ to vary over time, as shown in Fig. 7(a), where we indicate the dead-band and saturation frequencies using black dotted lines. Fig. 7(b) illustrates effective tracking of the required quantity by the delivered quantity, with the total cost aligning with the optimal value. Overall, the problem is solved within approximately $0.14$ s.

\section{Conclusions}
In this paper, we have proposed feedback-feedforward algorithms for distributed TVRA problems. In particular, we have introduced an initialization-free alternative to the barrier-based methods for handling LFCs, namely feedback laws based on projection and feedforward laws that switch between different modes. To govern the feedforward laws, we have devised a state-dependent switching signal that synchronizes with the projection operation, which reduces the number of switches without compromising optimality. We have conducted convergence analyses under mild assumptions: the proposed algorithm converges to the optimal trajectory within a fixed time for cases without LFCs, and is globally asymptotically stable at the optimal trajectory while exhibiting fixed-time convergence between switches for cases with LFCs. From a practical point of view, a key advantage of the projection-based method lies in its initialization-free capability coupled with a guarantee of local feasibility: should the state trajectory starts within the locally feasible set, it will remain therein; otherwise, it will enter the set within a fixed time and remain therein. Practically, it demonstrates global fixed-time convergence to the optimal trajectory and the effects of switching on convergence are negligible. The effectiveness of the proposed algorithms have been verified through numerical examples and a power system application. Future work includes extending the projection-based method to address global nonlinear inequality constraints.

\section{Appendix}
\subsection{Proof of Lemma \ref{lemma optimal trajectory}}
The Lagrangian of problem (\ref{problem}) is
\begin{align}\label{lagrangian}
    L =  \sum_{i=1}^N f_i(x_i,t) + \bar\lambda^\top\sum_{i=1}^N(A_ix_i-b_i(t)).
\end{align}
According to convex optimization theory \cite{boyd2004convex}, the optimal trajectory of problem (\ref{problem}) satisfy the KKT conditions:
\begin{subequations}\label{KKT conditions}
\begin{align}
    \mathbf 0_n &=  f_{ix}(x_i^\star,t) + A_i^\top\lambda^\star,\\
    \mathbf 0_l &= \sum_{i=1}^N(A_ix_i^\star - b_i(t)),
\end{align}
\end{subequations}
where $x_i^\star\in\mathbb R^{n}$ and $\lambda^\star\in\mathbb R^{l}$ denote the optimal trajectory for $x_i$ and $\bar\lambda$.  
Taking the time derivatives of both sides of (\ref{KKT conditions}a) and (\ref{KKT conditions}b) gives
\begin{subequations}\label{derivative optimality conditions}
\begin{align}
    \mathbf 0_n &= f_{ixx}(x_i^\star,t)\dot x_i^\star + f_{ixt}(x_i^\star,t) + A_i^\top\dot\lambda^\star,\\
    \mathbf 0_l &= \sum_{i=1}^N\left(A_i\dot x_i^\star - b_{it}(t)\right).
\end{align}    
\end{subequations}
Since $f_{i}(x_i,t)$ is uniformly strongly convex with respect to $x_i$, its inverse $f_{ixx}(x_i,t)^{-1}$ exists. Thus, (\ref{analytical expressions}a) follows directly from (\ref{derivative optimality conditions}a). Substituting (\ref{analytical expressions}a) and (\ref{weight}) into (\ref{derivative optimality conditions}b) yields (\ref{analytical expressions}b), which completes the proof.

\subsection{Proof of Lemma \ref{lemma distributed estimator}}
Denote the consensus error as $e_{\Delta_i} = \Delta_i - \frac{1}{N}\sum_{j=1}^N\Delta_j$. Consider the Lyapunov candidate $W_{\Delta} = \frac{1}{2}\sum_{i=1}^N \Vert e_{\Delta_i} \Vert_2^2$. Taking its time derivative, we obtain $\dot W_{\Delta}  = \sum_{i=1}^N e_{\Delta_i}^\top(\dot\Delta_i-\frac{1}{N}\sum_{j=1}^N\dot\Delta_j)$. Since $\sum_{i=1}^N e_{\Delta_i} = \mathbf 0_l$, it follows that $\dot W_{\Delta}  = \sum_{i=1}^Ne_{\Delta_i}^\top\dot\Delta_i$. We now decompose $\dot W_{\Delta}$ into three parts: $\dot W_{\Delta,1}$ corresponding to $(\cdot)^{1-\frac{p}{q}}$, $\dot W_{\Delta,2} $ corresponding to $(\cdot)^{1+\frac{p}{q}}$, and $\dot W_{\Delta,3}$ including the remaining terms. The first part can be obtained as
\begin{align}
    \dot W_{\Delta,1}
    &= -\gamma_1\sum_{i=1}^N\sum_{j=1}^Na_{ij} e_{\Delta_i}^\top(e_{\Delta_i}-e_{\Delta_j})^{1-\frac{p}{q}}\notag\\
    &= -\frac{\gamma_1}{2}\sum_{i=1}^N\sum_{j=1}^Na_{ij} e_{\Delta_i}^\top(e_{\Delta_i}-e_{\Delta_j} )^{1-\frac{p}{q}}\notag\\
    &\quad -\frac{\gamma_1}{2}\sum_{j=1}^N\sum_{i=1}^Na_{ji}e_{\Delta_j} ^\top(e_{\Delta_j}-e_{\Delta_i} )^{1-\frac{p}{q}}\notag\\
    &= -\frac{\gamma_1}{2}\sum_{i=1}^N\sum_{j=1}^Na_{ij}\Vert e_{\Delta_i} -e_{\Delta_j}  \Vert^{2-\frac{p}{q}}_{2-\frac{p}{q}}.\label{part 1}
\end{align}
For the second part, we have
\begin{align}
    \dot W_{\Delta,2}
    &= -\gamma_2\sum_{i=1}^N\sum_{j=1}^Na_{ij} e_{\Delta_i}^\top(e_{\Delta_i}-e_{\Delta_j})^{1+\frac{p}{q}}\notag\\
    &= -\frac{\gamma_2}{2}\sum_{i=1}^N\sum_{j=1}^Na_{ij}\Vert e_{\Delta_i} -e_{\Delta_j}  \Vert_{2+\frac{p}{q}}^{2+\frac{p}{q}}.\label{part 2}
\end{align}
With $\gamma_{3,\Delta}\geq\Vert A_i\dot x_i-b_{it}(t)\Vert_2$ $\forall t\geq 0$ we have
\begin{align}
    &\sum_{i=1}^N e_{\Delta_i}^\top(b_{it}(t)-A_i\dot x_i)\notag\\
    &=\frac{1}{N}\sum_{j=1}^N\sum_{i=1}^N e_{\Delta_i}^\top(b_{it}(t)-A_i\dot x_j)\notag\\
    &\leq \frac{1}{2N}\sum_{i=1}^N\sum_{j=1}^N\Vert e_{\Delta_i}-e_{\Delta_j}\Vert_2\Vert A_i\dot x_i-b_{it}(t)\Vert_2\notag\\
    &\leq \frac{\gamma_{3,\Delta}}{N(N-1)}\sum_{i=1}^N\sum_{j=1}^N \Vert e_{\Delta_i}-e_{\Delta_j}\Vert_2\notag\\
    &\leq^{(c)} \frac{\gamma_{3,\Delta}}{2}\sum_{i=1}^N\sum_{j=1}^N a_{ij}\Vert e_{\Delta_i}-e_{\Delta_j}\Vert_1,\notag
\end{align}
where $\leq^{(c)}$ is due to the facts that $\sum_{i=1}^N\sum_{j=1}^N \Vert e_{\Delta_i}-e_{\Delta_j} \Vert_2 \leq N\max_{i\in\mathcal N}\sum_{j=1,j\neq i}^N\Vert e_{\Delta_i}-e_{\Delta_j}\Vert_2 \leq \frac{N(N-1)}{2}\sum_{i=1}^N\sum_{j=1}^N a_{ij}\Vert e_{\Delta_i}-e_{\Delta_j} \Vert_2$ and $\Vert\cdot\Vert_2\leq\Vert\cdot\Vert_1$. Thus, the third part can be obtained as
\begin{align}\label{part 3}
    \dot W_{\Delta,3}
    &\leq \frac{\gamma_{3,\Delta}-\gamma_{3,\Delta}}{2}\sum_{i=1}^N\sum_{j=1}^Na_{ij}\Vert e_{\Delta_i}-e_{\Delta_j} \Vert_1 = 0.
\end{align}
For unweighted graphs, the combination of (\ref{part 1})--(\ref{part 3}) leads to 
\begin{align}
    \dot W_{\Delta} 
    &\leq -\frac{\gamma_1}{2}\left[\sum_{i=1}^N\sum_{j=1}^Na_{ij}^\frac{2q}{2q-p}\Vert e_{\Delta_i}-e_{\Delta_j}  \Vert_2^2\right]^\frac{2q-p}{2q}\notag\\
    &\quad - \frac{\gamma_2}{2}l^{-\frac{p}{2q}}N^{-\frac{p}{q}}\left[\sum_{i=1}^N\sum_{j=1}^Na_{ij}^\frac{2q}{2q+p}\Vert e_{\Delta_i}-e_{\Delta_j}\Vert_2^2\right]^\frac{2q+p}{2q}\notag\\
    &= -\frac{\gamma_1}{2}\left[2e_{\Delta}^\top(\mathcal L\otimes I_n)e_{\Delta}\right]^{1-\frac{p}{2q}}\notag\\
    &\quad - \frac{\gamma_2}{2}l^{-\frac{p}{2q}}N^{-\frac{p}{q}}\left[2e_{\Delta}^\top(\mathcal L\otimes I_n)e_{\Delta}\right]^{1+\frac{p}{2q}}\notag\\
    &\leq -\frac{\gamma_1}{2}\left[2\eta_2(\mathcal L)e_{\Delta}^\top e_{\Delta}\right]^{1-\frac{p}{2q}}\notag\\
    &\quad - \frac{\gamma_2}{2}l^{-\frac{p}{2q}}N^{-\frac{p}{q}}\left[2\eta_2(\mathcal L)e_{\Delta}^\top e_{\Delta}\right]^{1+\frac{p}{2q}}\notag\\
    &= -\frac{\gamma_1}{2}\left[4\eta_2(\mathcal L)W_{\Delta}\right]^{1-\frac{p}{2q}}\notag\\
    &\quad - \frac{\gamma_2}{2}l^{-\frac{p}{2q}}N^{-\frac{p}{q}}\left[4\eta_2(\mathcal L)W_{\Delta}\right]^{1+\frac{p}{2q}}.
\end{align}

By Lemma 1, $e_{\Delta_i}$ is fixed-time stable at the origin, which means $\Delta_i=\Delta_j$ $\forall t\geq T_1$ $\forall i,j\in\mathcal N$. Meanwhile,
\begin{align}
    \sum_{i=1}^N\Delta_i(t) &= \sum_{i=1}^N\zeta_i(t) + \sum_{i=1}^N(b_i(t)-A_ix_i)\notag\\
    &= \sum_{i=1}^N\zeta_i(0) + \int_{0}^{t}\sum_{i=1}^N\dot\zeta_i(t)dt + \sum_{i=1}^N(b_i(t)-A_ix_i)\notag\\
    &= \sum_{i=1}^N(b_i(t)-A_ix_i),
\end{align}
which gives rise to $\Delta_i \to -\frac{1}{N}\sum_{i=1}^N(A_ix_i-b_i(t))$ for all $i\in\mathcal N$. Given the symmetry between (\ref{distributed estimator}b)--(\ref{distributed estimator}c), (\ref{distributed estimator}d)--(\ref{distributed estimator}e), and (\ref{distributed estimator}f)--(\ref{distributed estimator}g), their analyses follow similar steps. With $\gamma_{3,\psi}\geq\frac{N-1}{2}\Vert\dot\rho_i\Vert_2$ and $\gamma_{3,\psi'}\geq\frac{N-1}{2}\Vert\dot\phi_i\Vert_2$, it is not difficult to show $\psi_i=\psi_j$ and $\psi'_i=\psi'_j$ $\forall t\geq T_1^{\max}$ $\forall i,j\in\mathcal N$, and $y_i \to -\left[\sum_{i=1}^N\rho_i\right]^{-1}\sum_{i=1}^N\phi_i$ $\forall t\geq T_1^{\max}$ $\forall i\in\mathcal N$.

\subsection{Non-Zenoness of Switching}
Consider a compact set $[0,T]$. Since $x_i$ is absolutely continuous, $F_{i,x}$ is uniformly continuous on $[0,T]$, and the projection operator $\mathcal P_{X_i(t)}$ is 1-Lipschitz, $x_i-e_i = \mathcal P_{X_i(t)}(x_i-F_{i,x})$ and $g_{i,m}(x_i-e_i,t)$ are uniformly continuous on $[0,T]$. In what follows, we fix an arbitrary $i_0\in\mathcal N$ and $m_0\in\mathcal M$. For notational simplicity, define $\xi(t) = g_{i_0,m_0}(x_{i_0}-e_{i_0},t)$, which is differentiable almost everywhere, and $M= \max_{t\in[0,T]}\vert\dot\xi(t)\vert$. Regarding three consecutive switching instants where $\sigma_{i_0,m_0}$ undergoes transitions from ``$0$'' to ``$1$'', then ``$1$'' to ``$0$'', and finally ``$0$'' to ``$1$'', we have $\xi(t_{k-1})=\xi(t_{k+1})=0$ and $\xi(t_{k})<0$. If we define
\begin{align}
    s_k=\mathrm{arg}\max\nolimits_{t\in(t_{k-1},t_{k+1})}\vert\xi(t)\vert,
\end{align}
it follows by the mean value theorem that there is a point $r_k\in(t_{k-1},s_k)$ where
\begin{align}
    \vert\dot\xi(r_k)\vert = \frac{\vert\xi(s_k)-0\vert}{s_k-t_{k-1}}\leq M.
\end{align}
Likewise, there is a point $q_k\in(s_k,t_{k+1})$ where
\begin{align}
    \vert\dot\xi(q_k)\vert = \frac{\vert 0-\xi(s_k)\vert}{t_{k+1}-s_k} \leq M.
\end{align}
Hence, we have
\begin{align}
    t_{k+1}-t_{k-1} = t_{k+1}-s_k + s_k-t_{k-1} \geq \frac{2\vert\xi(s_k)\vert}{M}.
\end{align}
For the switching to be non-Zeno, it remains to show that $\vert\xi(s_k)\vert$ does not vanish as $k\to+\infty$.

Since $\xi(t)$ is uniformly continuous on $[0,T]$, for any $\epsilon>0$, there always exists $\delta>0$ such that
\begin{align}\label{here we can substitute}
    \vert\xi(s) - \xi(t)\vert < \epsilon
\end{align}
for all $s,t$ in $[0,T]$ satisfying $\vert s-t\vert < \delta$. Suppose for contradiction that there are infinitely many switches within $T$. This means $t_{k+1}-t_{k-1}\to 0$ and $\xi(s_k)\to 0$ as $k\to+\infty$, and we must have $s_{k}-t_{k-1}<t_{k+1}-t_{k-1}<\delta$ for some sufficiently small $\delta>0$. Accordingly, we can proceed by substituting $s=s_k$, $t=t_{k-1}$, and $\epsilon = \vert\xi(s_k)\vert$ into (\ref{here we can substitute}):
\begin{align}
    \vert\xi(s_k) - \xi(t_{k-1})\vert < \vert\xi(s_k)\vert,\ k\to+\infty,
\end{align}
which is a contradiction to the fact that
\begin{align}
    \vert\xi(s_k) - \xi(t_{k-1})\vert = \vert\xi(s_k)\vert,\ \forall k.
\end{align}
Thus, there should be only finitely many switches within $T$, and the minimum two-step dwell time is described by
\begin{align}
    \tau\geq\frac{2\min\nolimits_{k}\vert\xi(s_k)\vert}{M}>0,
\end{align}
where
\begin{align}
    \vert\xi(s_k)\vert &= \max\nolimits_{t\in(t_k,t_{k+1})}\vert g_{i_0,m_0}(\mathcal P_{X_{i_0}(t)}(x_{i_0}-F_{i_0,x}),t)\vert\notag\\
    &\geq\max\nolimits_{t\in(t_k,t_{k+1})}\vert g_{i_0,m_0}(\mathcal P_{X_{i_0}(t)}(x_{i_0}),t)\vert.\label{enlarge}
\end{align}
Enlarging the two-step dwell time via (\ref{enlarge}), the enhanced switching rule contributes to a reduction in the number of switches needed to reach the optimal trajectory.

\section*{References}
\bibliographystyle{ieeetr}
\bibliography{ref}

\begin{thebibliography}{10}

\bibitem{7469393}
A.~Simonetto, A.~Mokhtari, A.~Koppel, G.~Leus, and A.~Ribeiro, ``A class of prediction-correction methods for time-varying convex optimization,'' {\em IEEE Transactions on Signal Processing}, vol.~64, no.~17, pp.~4576--4591, 2016.

\bibitem{8782575}
X.~Xing, J.~Lin, N.~Brandon, A.~Banerjee, and Y.~Song, ``Time-varying model predictive control of a reversible-soc energy-storage plant based on the linear parameter-varying method,'' {\em IEEE Transactions on Sustainable Energy}, vol.~11, no.~3, pp.~1589--1600, 2020.

\bibitem{9810516}
B.~Huang, Y.~Zou, F.~Chen, and Z.~Meng, ``Distributed time-varying economic dispatch via a prediction-correction method,'' {\em IEEE Transactions on Circuits and Systems I: Regular Papers}, vol.~69, no.~10, pp.~4215--4224, 2022.

\bibitem{9133310}
A.~Simonetto, E.~Dall'Anese, S.~Paternain, G.~Leus, and G.~B. Giannakis, ``Time-varying convex optimization: Time-structured algorithms and applications,'' {\em Proceedings of the IEEE}, vol.~108, no.~11, pp.~2032--2048, 2020.

\bibitem{su2009traffic}
W.~Su, {\em Traffic engineering and time-varying convex optimization}.
\newblock The Pennsylvania State University, 2009.

\bibitem{9855238}
S.~Sun, J.~Xu, and W.~Ren, ``Distributed continuous-time algorithms for time-varying constrained convex optimization,'' {\em IEEE Transactions on Automatic Control}, vol.~68, no.~7, pp.~3931--3946, 2023.

\bibitem{9670443}
B.~Wang, S.~Sun, and W.~Ren, ``Distributed time-varying quadratic optimal resource allocation subject to nonidentical time-varying hessians with application to multiquadrotor hose transportation,'' {\em IEEE Transactions on Systems, Man, and Cybernetics: Systems}, vol.~52, no.~10, pp.~6109--6119, 2022.

\bibitem{7993088}
A.~Simonetto and E.~Dall’Anese, ``Prediction-correction algorithms for time-varying constrained optimization,'' {\em IEEE Transactions on Signal Processing}, vol.~65, no.~20, pp.~5481--5494, 2017.

\bibitem{8502778}
A.~Simonetto, ``Dual prediction–correction methods for linearly constrained time-varying convex programs,'' {\em IEEE Transactions on Automatic Control}, vol.~64, no.~8, pp.~3355--3361, 2019.

\bibitem{8062794}
M.~Fazlyab, S.~Paternain, V.~M. Preciado, and A.~Ribeiro, ``Prediction-correction interior-point method for time-varying convex optimization,'' {\em IEEE Transactions on Automatic Control}, vol.~63, no.~7, pp.~1973--1986, 2018.

\bibitem{7902101}
A.~Simonetto, A.~Koppel, A.~Mokhtari, G.~Leus, and A.~Ribeiro, ``Decentralized prediction-correction methods for networked time-varying convex optimization,'' {\em IEEE Transactions on Automatic Control}, vol.~62, no.~11, pp.~5724--5738, 2017.

\bibitem{7862771}
C.~Sun, M.~Ye, and G.~Hu, ``Distributed time-varying quadratic optimization for multiple agents under undirected graphs,'' {\em IEEE Transactions on Automatic Control}, vol.~62, no.~7, pp.~3687--3694, 2017.

\bibitem{8100702}
B.~Ning, Q.-L. Han, and Z.~Zuo, ``Distributed optimization for multiagent systems: An edge-based fixed-time consensus approach,'' {\em IEEE Transactions on Cybernetics}, vol.~49, no.~1, pp.~122--132, 2019.

\bibitem{7518617}
S.~Rahili and W.~Ren, ``Distributed continuous-time convex optimization with time-varying cost functions,'' {\em IEEE Transactions on Automatic Control}, vol.~62, no.~4, pp.~1590--1605, 2017.

\bibitem{9629369}
H.~Hong, S.~Baldi, W.~Yu, and X.~Yu, ``Distributed time-varying optimization of second-order multiagent systems under limited interaction ranges,'' {\em IEEE Transactions on Cybernetics}, vol.~52, no.~12, pp.~13874--13886, 2022.

\bibitem{chen2023time}
Y.~Chen, T.~Yu, Q.~Meng, F.~Niu, and H.~Wang, ``Time-varying distributed optimization problem with inequality constraints,'' {\em Journal of the Franklin Institute}, vol.~360, no.~16, pp.~11314--11330, 2023.

\bibitem{li2020time}
Z.~Li and Z.~Ding, ``Time-varying multi-objective optimisation over switching graphs via fixed-time consensus algorithms,'' {\em Int. J. Syst. Sci.}, vol.~51, no.~15, pp.~2793--2806, 2020.

\bibitem{9827560}
C.~Wu, H.~Fang, X.~Zeng, Q.~Yang, Y.~Wei, and J.~Chen, ``Distributed continuous-time algorithm for time-varying optimization with affine formation constraints,'' {\em IEEE Transactions on Automatic Control}, vol.~68, no.~4, pp.~2615--2622, 2023.

\bibitem{zhao2025distributed}
X.~Zhao, H.~Wu, and J.~Cao, ``Distributed consensus time-varying optimization algorithm for multi-agent systems in predefined time via event-triggered sliding mode control,'' {\em Mathematical Methods in the Applied Sciences}, vol.~48, no.~2, pp.~1617--1635, 2025.

\bibitem{9183902}
B.~Wang, S.~Sun, and W.~Ren, ``Distributed continuous-time algorithms for optimal resource allocation with time-varying quadratic cost functions,'' {\em IEEE Transactions on Control of Network Systems}, vol.~7, no.~4, pp.~1974--1984, 2020.

\bibitem{8619295}
L.~Bai, C.~Sun, Z.~Feng, and G.~Hu, ``Distributed continuous-time resource allocation with time-varying resources under quadratic cost functions,'' in {\em 2018 IEEE Conference on Decision and Control (CDC)}, pp.~823--828, 2018.

\bibitem{9123943}
B.~Wang, Q.~Fei, and Q.~Wu, ``Distributed time-varying resource allocation optimization based on finite-time consensus approach,'' {\em IEEE Control Systems Letters}, vol.~5, no.~2, pp.~599--604, 2021.

\bibitem{10239516}
X.~Shi, G.~Wen, and X.~Yu, ``Finite-time convergent algorithms for time-varying distributed optimization,'' {\em IEEE Control Systems Letters}, vol.~7, pp.~3223--3228, 2023.

\bibitem{10791871}
S.~Jiang and Z.~Ding, ``Distributed optimal resource allocation control for heterogeneous linear multiagent systems,'' {\em IEEE Transactions on Automatic Control}, pp.~1--8, 2024.

\bibitem{9953144}
Z.~Ding, ``Distributed time-varying optimization—an output regulation approach,'' {\em IEEE Transactions on Cybernetics}, pp.~1--11, 2022.

\bibitem{isidori1990output}
A.~Isidori and C.~I. Byrnes, ``Output regulation of nonlinear systems,'' {\em IEEE transactions on Automatic Control}, vol.~35, no.~2, pp.~131--140, 1990.

\bibitem{9893893}
H.~Hong, W.~Yu, G.-P. Jiang, and H.~Wang, ``Fixed-time algorithms for time-varying convex optimization,'' {\em IEEE Transactions on Circuits and Systems II: Express Briefs}, vol.~70, no.~2, pp.~616--620, 2023.

\bibitem{10829801}
G.~Guo, Z.-D. Zhou, and R.~Zhang, ``Distributed time-varying constrained convex optimization: Finite-/fixed-time convergence,'' {\em IEEE Transactions on Control of Network Systems}, pp.~1--12, 2025.

\bibitem{8746216}
I.~Notarnicola and G.~Notarstefano, ``Constraint-coupled distributed optimization: A relaxation and duality approach,'' {\em IEEE Transactions on Control of Network Systems}, vol.~7, no.~1, pp.~483--492, 2020.

\bibitem{boyd2004convex}
S.~Boyd, ``Convex optimization,'' {\em Cambridge UP}, 2004.

\bibitem{9762539}
X.~Wu, H.~Wang, and J.~Lu, ``Distributed optimization with coupling constraints,'' {\em IEEE Transactions on Automatic Control}, vol.~68, no.~3, pp.~1847--1854, 2023.

\bibitem{huang2024distributed}
Y.~Huang, Z.~Meng, J.~Sun, and G.~Wang, ``Distributed continuous-time proximal algorithm for nonsmooth resource allocation problem with coupled constraints,'' {\em Automatica}, vol.~159, p.~111309, 2024.

\bibitem{9141512}
G.~Chen and Z.~Guo, ``Initialization-free distributed fixed-time convergent algorithms for optimal resource allocation,'' {\em IEEE Transactions on Systems, Man, and Cybernetics: Systems}, vol.~52, no.~2, pp.~845--854, 2022.

\bibitem{yi2016initialization}
P.~Yi, Y.~Hong, and F.~Liu, ``Initialization-free distributed algorithms for optimal resource allocation with feasibility constraints and application to economic dispatch of power systems,'' {\em Automatica}, vol.~74, pp.~259--269, 2016.

\bibitem{liu2024achieving}
C.~Liu, X.~Tan, X.~Wu, D.~V. Dimarogonas, and K.~H. Johansson, ``Achieving violation-free distributed optimization under coupling constraints,'' {\em arXiv preprint arXiv:2404.07609}, 2024.

\bibitem{10909192}
X.~Yi, X.~Li, T.~Yang, L.~Xie, Y.~Hong, T.~Chai, and K.~H. Johansson, ``Distributed online convex optimization with time-varying constraints: Tighter cumulative constraint violation bounds under slater's condition,'' {\em IEEE Transactions on Automatic Control}, pp.~1--16, 2025.

\bibitem{10876582}
X.~Tan, C.~Liu, K.~H. Johansson, and D.~V. Dimarogonas, ``A continuous-time violation-free multi-agent optimization algorithm and its applications to safe distributed control,'' {\em IEEE Transactions on Automatic Control}, pp.~1--15, 2025.

\bibitem{8322314}
Z.~Zuo, Q.-L. Han, B.~Ning, X.~Ge, and X.-M. Zhang, ``An overview of recent advances in fixed-time cooperative control of multiagent systems,'' {\em IEEE Transactions on Industrial Informatics}, vol.~14, no.~6, pp.~2322--2334, 2018.

\bibitem{shevitz1994lyapunov}
D.~Shevitz and B.~Paden, ``Lyapunov stability theory of nonsmooth systems,'' {\em IEEE Transactions on Automatic Control}, vol.~39, no.~9, pp.~1910--1914, 1994.

\bibitem{fukushima1992equivalent}
M.~Fukushima, ``Equivalent differentiable optimization problems and descent methods for asymmetric variational inequality problems,'' {\em Mathematical programming}, vol.~53, pp.~99--110, 1992.

\bibitem{9796617}
Y.~Gu and T.~C. Green, ``Power system stability with a high penetration of inverter-based resources,'' {\em Proceedings of the IEEE}, vol.~111, no.~7, pp.~832--853, 2023.

\bibitem{NGESO_guide}
\text{National Grid ESO}, ``New dynamic response services: Provider guidance v.8.'' [Online], 2024.
\newblock \url{https://www.nationalgrideso.com/document/276606/download}.

\bibitem{shi2018analytical}
Q.~Shi, F.~Li, and H.~Cui, ``Analytical method to aggregate multi-machine sfr model with applications in power system dynamic studies,'' {\em IEEE Transactions on Power Systems}, vol.~33, no.~6, pp.~6355--6367, 2018.

\bibitem{kundur}
P.~S. Kundur, {\em Power System Stability and Control}.
\newblock New York, NY, USA: McGraw-Hill Education, 1994.

\end{thebibliography}

\begin{IEEEbiography}[{\includegraphics[width=1in,height=1.25in,clip,keepaspectratio]{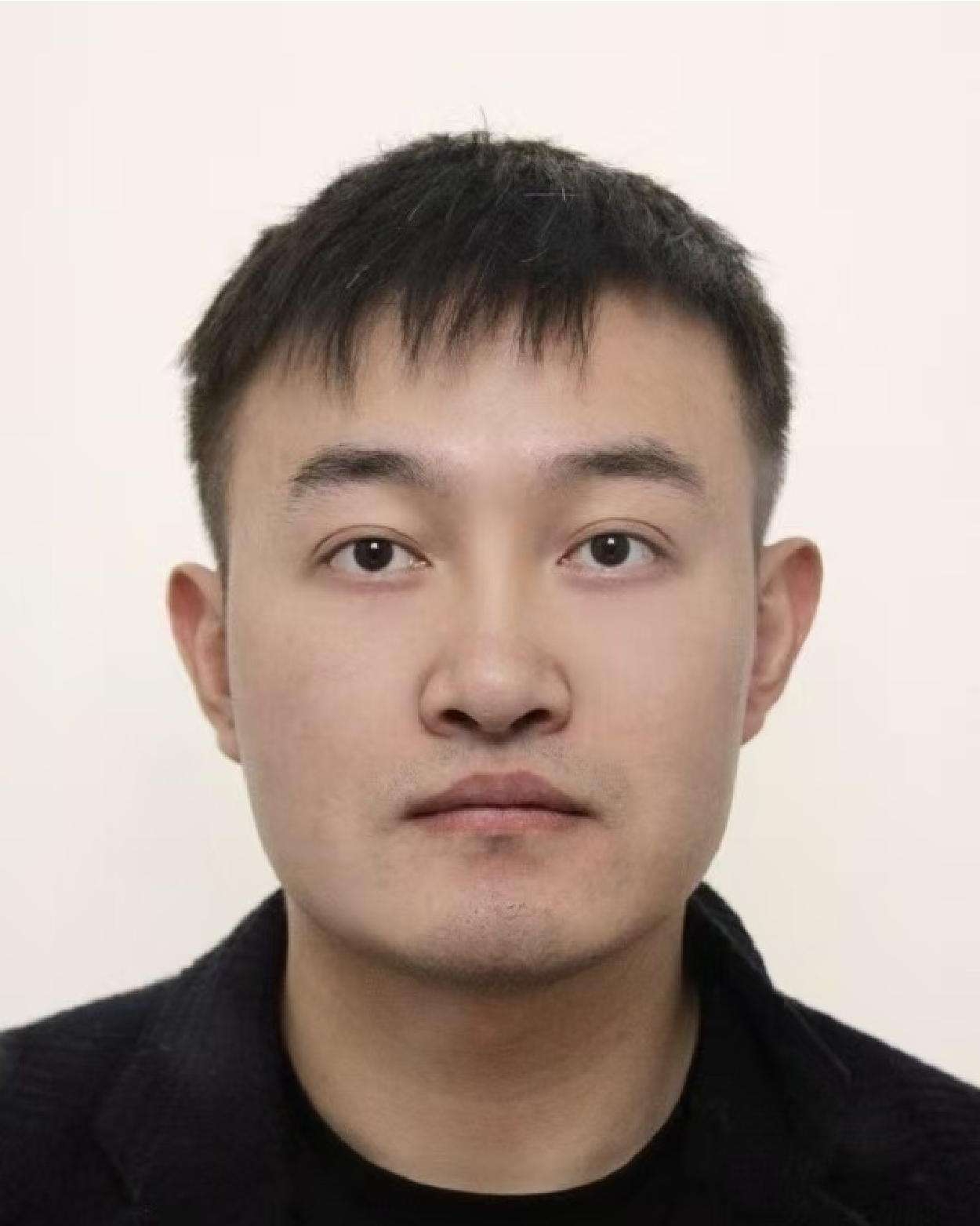}}]%
{Yiqiao Xu} received the M.Sc. degree in Advanced Control and Systems Engineering in 2018 and the Ph.D. degree in Electrical and Electronic Engineering in 2023 from The University of Manchester, U.K. Since 2023, he has been a postdoctoral research associate in the Power and Energy Division at the same institution. His research interests include distributed optimization and learning-based control of power networks and multi-energy systems.
\end{IEEEbiography}

\begin{IEEEbiography}[{\includegraphics[width=1in,height=1.25in,clip,keepaspectratio]{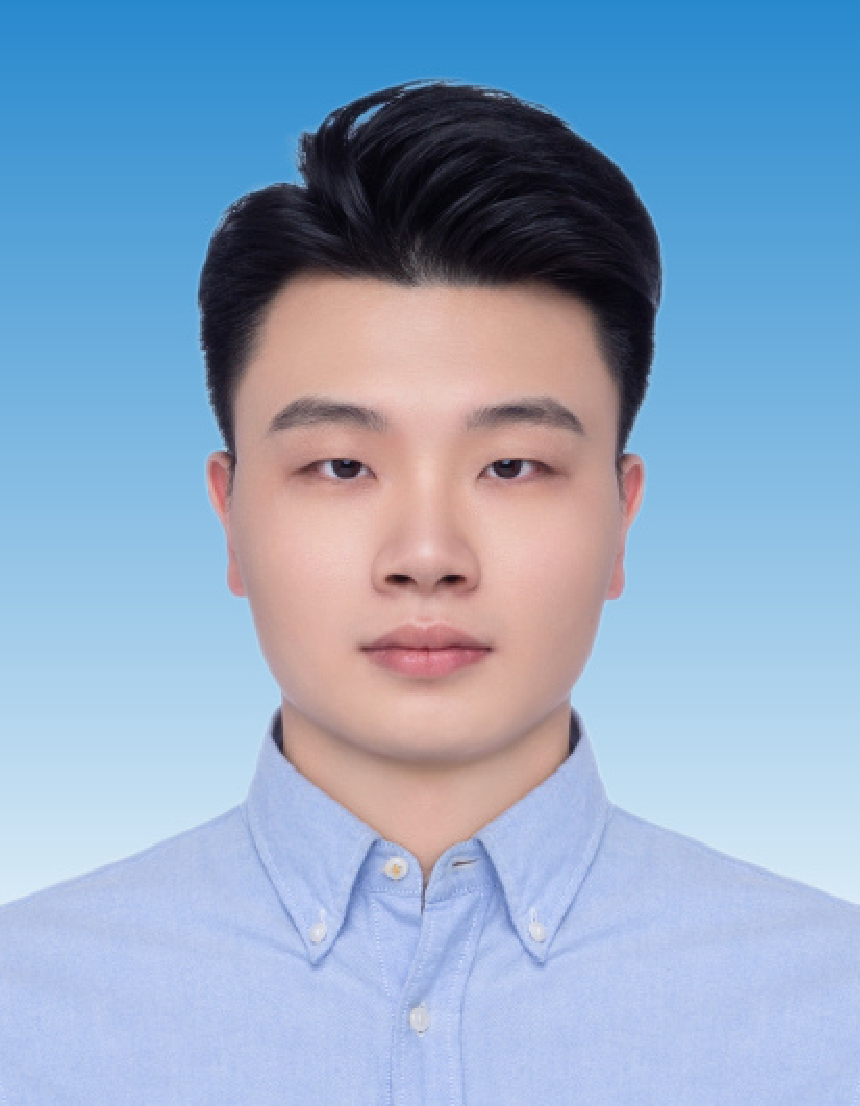}}]%
{Tengyang Gong} received the M.Sc. degree from the Department of Electrical and Electronic Engineering at The University of Manchester, U.K., in 2022. He is currently pursuing a Ph.D. in the same department. His research interests include the distributed optimization, control, and game theory of multi-agent systems. He was awarded the Best Theory Paper Award at the IEEE 30th International Conference on Automation and Computing (ICAC 2025).
\end{IEEEbiography}

\begin{IEEEbiography}[{\includegraphics[width=1in,height=1.25in,clip,keepaspectratio]{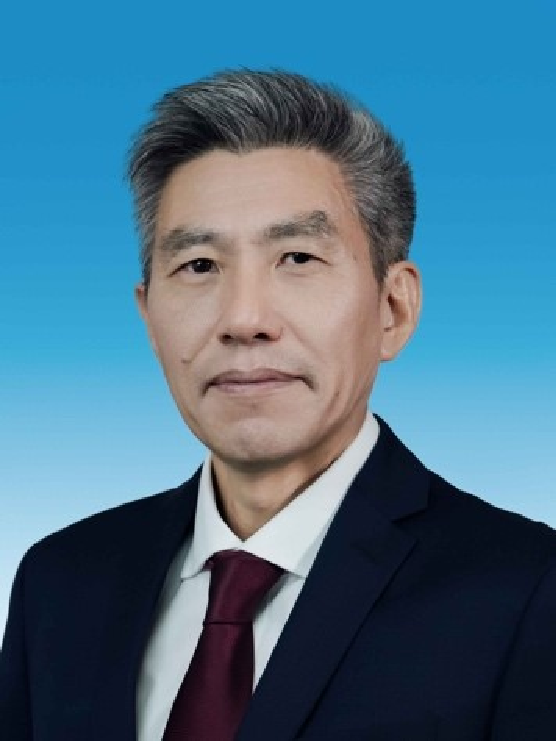}}]%
{Zhengtao Ding} received the B.Eng. degree from Tsinghua University, Beijing, China, in 1984, and the M.Sc. degree in systems and control and the Ph.D. degree in control systems from the University of Manchester Institute of Science and Technology, Manchester, U.K., in 1986 and 1989, respectively. He is a Professor of Control Systems with the Department of Electrical and Electronic Engineering. He is the author of the book Nonlinear and Adaptive Control Systems (IET, 2013) and has published over 300 research papers. His research interests include nonlinear and adaptive control theory and their applications, more recently, network-based control, distributed optimization, and distributed machine learning, with applications to power systems and robotics. Prof. Ding has served as an Associate Editor for the IEEE Transactions on Automatic Control, IEEE Control Systems Letters, and several other journals. He is a member of the IEEE Technical Committee on Nonlinear Systems and Control, IEEE Technical Committee on Intelligent Control, and IFAC Technical Committee on Adaptive and Learning Systems. He is a Fellow of The Alan Turing Institute, the U.K. national institute for data science and artificial intelligence.
\end{IEEEbiography}

\begin{IEEEbiography}[{\includegraphics[width=1in,height=1.25in,clip,keepaspectratio]{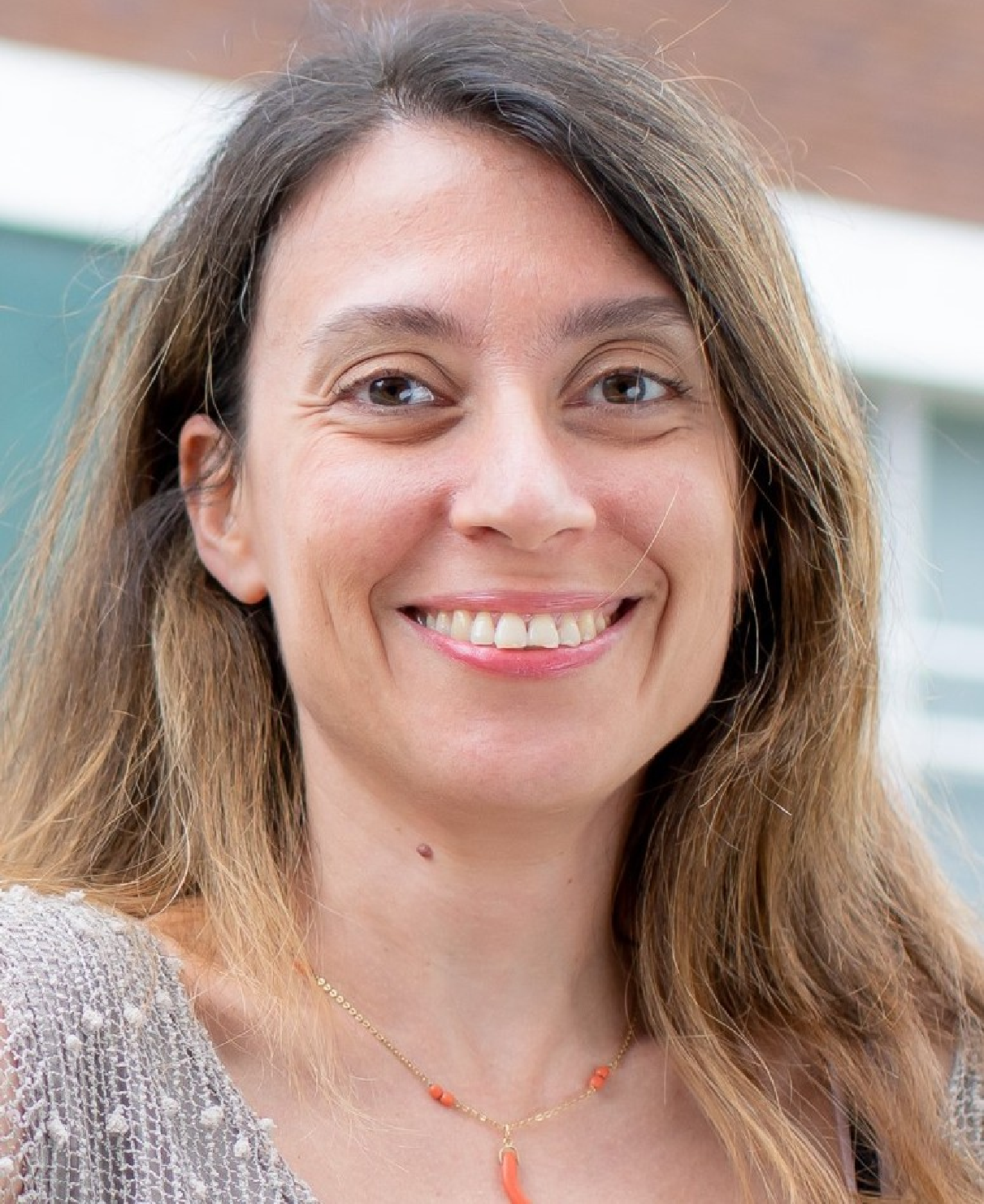}}]%
{Alessandra Parisio} is a Professor of Control of Sustainable Energy Networks, in the Department of Electrical and Electronic Engineering at The University of Manchester, U.K. She is IEEE senior member, co-chair of the IEEE RAS Technical Committee on Smart Buildings and vice-Chair for Education of the IFAC Technical Committee 9.3. Control for Smart Cities. She is currently an Associate Editor of the IEEE Transactions on Control of Network Systems, IEEE Transactions on Automation Science and Engineering, European Journal of Control and Applied Energy. Her main research interests span the areas of control engineering, in particular Model Predictive Control, distributed optimization and control, stochastic constrained control, and power systems, with energy management systems under uncertainty, optimization and control of multi-energy networks and distributed flexibility.
\end{IEEEbiography}

\end{document}